\documentclass[a4paper,11pt]{article}
\usepackage{jcappub}
\usepackage{graphicx} 
\usepackage[utf8]{inputenc}
\usepackage{physics}
\usepackage{amsfonts,amssymb,amsmath}
\usepackage{mathtools}

\usepackage{xcolor}

\usepackage[acronym, toc, nonumberlist]{glossaries}

\glsdisablehyper
\setacronymstyle{long-short}

\newacronym{ns}{NS}{neutron star}
\newacronym{bh}{BH}{black hole}
\newacronym{bbh}{BBH}{binary black hole}
\newacronym{bns}{BNS}{binary neutron star}
\newacronym{nsbh}{NSBH}{neutron star black hole}

\newacronym{eos}{EoS}{equation of state}
\newacronym{gw}{GW}{gravitational wave}
\newacronym{gr}{GR}{general relativity}
\newacronym{snr}{SNR}{signal-to-noise ratio}

\newacronym{lisa}{LISA}{Laser Interferometer Space Antenna }
\newacronym{ligo}{LIGO}{Laser Interferometer Gravitational wave Observatory}
\newacronym{kagra}{KAGRA}{KAmioka GRavitational wave detector}
\newacronym{eob}{EOB}{effective one-body}
\newacronym{em}{EM}{electromagnetic}
\newacronym{lcdm}{$\Lambda$CDM}{$\Lambda$ cold dark matter}
\newacronym{pl}{PL}{power law}
\newacronym{plg}{PLG}{power law and Gaussian}
\newacronym{kde}{KDE}{kernel density estimate}
\newacronym{de}{DE}{dark energy}
\newacronym{cdf}{CDF}{cumulative density function}

\newacronym{lvk}{LVK}{LIGO-Virgo-KAGRA}
\newacronym{ego}{EGO}{European gravitational observatory}
\newacronym{asd}{ASD}{amplitude spectral density}
\newacronym{psd}{PSD}{power spectral density}
\newacronym{mcmc}{MCMC}{Monte Carlo Markov chain}
\newacronym{hlv}{HLV}{Hanford Livingston Virgo}
\newacronym{pe}{PE}{parameter estimation}
\newacronym{cbc}{CBC}{compact binary coalescence}
\newacronym{aligo}{aLIGO}{advanced LIGO}
\newacronym{far}{FAR}{false alarm rate}

\newacronym{cl}{CL}{confidence level}

\newacronym{pn}{PN}{post-Newtonian}
\newacronym{nr}{NR}{numerical relativity}
\newacronym{ppisn}{PPISN}{pulsation pair-instability supernova}
\newacronym{pisn}{PISN}{pair instability-supernova}
\newacronym{et}{ET}{Einstein Telescope}
\newacronym{ce}{CE}{Cosmic Explorer}

\newacronym{cmb}{CMB}{cosmic microwave background}
\newacronym{lss}{LSS}{large scale structure}
\newacronym{isco}{ISCO}{innermost stable orbit}

\newacronym{oi}{Oi}{observation run $i$}
\newacronym{gwtci}{GWTC-i}{gravitational wave transient catalog $i$}

\newacronym{2g}{2G}{second generation}
\newacronym{3g}{3G}{third generation}

\newacronym{nf}{NF}{normalizing flow}
\newacronym{ml}{ML}{machine learning}
\newacronym{lfi}{LFI}{likelihood-free inference}
\newacronym{nn}{NN}{neural network}
\newacronym{dingo}{DINGO}{deep inference for gravitational wave observations}
\newacronym{gpu}{GPU}{graphics processing unit}
\newacronym{hba}{HBA}{hierarchical Bayesian analysis}

\newacronym{kl}{KL}{Kullback-Leibler}
\newacronym{js}{JS}{Jensen-Shannon}
\newacronym{ks}{KS}{Kolmogorov–Smirnov}

\newacronym{dm}{DM}{dark matter}

\newcommand{\md}{m_{\rm d}}
\newcommand{\ms}{m_{\rm s}}

\newcommand{\mmax}{m_{\rm max}}

\newcommand{\msone}{m_{1,{\rm s}}}
\newcommand{\mstwo}{m_{2,{\rm s}}}

\newcommand{\data}{\mathcal{D}}
\newcommand{\datagw}{\data_{{\rm GW}}}
\newcommand{\datagwsub}[1]{\ensuremath{\data_{{\rm GW}, #1}}}

\newcommand{\setdgw}{\{\datagw\}}

\newcommand{\hyper}{\Lambda}
\newcommand{\param}{\lambda}
\newcommand{\params}{\boldsymbol{\param}}
\newcommand{\paramsset}{\{\param_{1},\param_{2},...\}}

\newcommand{\pdet}{p_{\rm det}(\params)}

\newcommand{\nobs}{N_{\rm obs}}

\newcommand{\model}{\mathfrak{M}}

\newcommand{\ninjtot}{N_{\rm inj, tot}}
\newcommand{\ninjsel}{N_{\rm inj, sel}}
\newcommand{\injparams}{\params_{\rm inj}}

\title{The impact of precession and higher-order multipoles for gravitational wave cosmological inference}
\author[a]{Charlie Hoy,}
\emailAdd{charlie.hoy@port.ac.uk}
\affiliation[a]{University of Portsmouth, Portsmouth, PO1 3FX, United Kingdom}
\author[a,b]{Konstantin Leyde}
\emailAdd{kleyde@flatironinstitute.org}
\affiliation[b]{Center for Computational Astrophysics, Flatiron Institute, 162 Fifth Avenue, New York, NY 10010, USA}
\date{\today}

\abstract{
Gravitational-wave astronomy presents an exciting opportunity to provide an independent measurement of the expansion rate of the Universe. 
By combining inferences for the binary component masses and luminosity distances from individual observations, it is possible to infer $H_0$ without direct electromagnetic counterparts or galaxy catalogs. 
However, this relies on theoretical gravitational-wave models, which are known to be incomplete descriptions of the full predictions of general relativity. Although the accuracy of our models are improving, they are also becoming increasingly expensive as additional phenomena are incorporated. In this work, we demonstrate that there is no significant advantage in including spin-precession and higher-order multipole moments when inferring $H_0$ via the mass spectrum method for current and near-future gravitational-wave event numbers. Even when simulating a population of highly precessing and preferentially asymmetric-mass-ratio binaries, we show that the inferred $H_0$ posterior changes minimally. Using a simpler, less accurate model, achieves comparable $H_0$ estimates with six times less computational cost (on average). Using computationally cheaper models for single event inference may become crucial as thousands of gravitational-wave observations are expected to be detected in the near future.
}

\begin{document}
\maketitle
\flushbottom

\section{Introduction}

\Gls{gw} astronomy has attracted significant attention as it provides an independent probe to measure the expansion rate of the Universe: $H_0$ \cite{LIGOScientific:2021aug, LIGOScientific:2025jau}. 
Indeed, the combined detection of \glspl{gw} from two neutron stars \cite{LIGOScientific:2017vwq} and its associated \gls{em} counterpart \cite{LIGOScientific:2017adf} allowed $H_0$ to be estimated within a relative uncertainty of $20\%$, broadly consistent with measurements from the cosmic microwave background \cite{Planck:2018vyg} and supernovae standard candles \cite{Riess:2019cxk, Freedman:2024eph}. 
An alternative method to estimate $H_0$ from \glspl{gw} uses the (redshifted) source-frame mass distribution, developed in \cite{Taylor:2011fs, Taylor:2012db} and further explored in \cite{Farr:2019twy, Mastrogiovanni:2021wsd, Mancarella:2021ecn, Mukherjee:2021rtw, Leyde:2022orh, Karathanasis:2022rtr, Ezquiaga:2022zkx, Pierra:2023deu, Leyde:2023iof, Borghi:2023opd, Farah:2024xub, MaganaHernandez:2024uty, Mali:2024wpq, Agarwal:2024hld, Mancarella:2025uat}. 
When the GW propagates on an expanding spacetime, the signal depends on the source-frame mass, $\ms$ through the combination of $\md = \ms (1 + z)$, where this new quantity is referred to as the detector-frame mass. 
The so-called \textit{mass spectrum} method, also referred to as the \textit{spectral siren} method, estimates $H_0$ by combining the measured detector-frame mass and the source-frame mass distribution into a redshift measurement. 
Using luminosity distance directly inferred from the \gls{gw} data, constraints can be placed on $H_0$ through the luminosity distance-redshift relation.
Other approaches include using galaxy catalogs to build a three-dimensional prior on the \gls{cbc} distribution \cite{Schutz:1986gp, DelPozzo:2011vcw, Gray:2019ksv, Gray:2021sew, Finke:2021aom, Mukherjee:2022afz, Turski:2023lxq, Mastrogiovanni:2023emh, Gray:2023wgj, Borghi:2023opd, Rauf:2023oiy, Perna:2024lod, Hanselman:2024hqy, Leyde:2024tov, Li:2025hrh, Naveed:2025kgk, Leyde:2025rzk}, often referred to as the \textit{dark siren} method. However, currently employed catalogs are essentially empty at the higher-end of luminosity distances associated with \glspl{gw}. This is why we do not see a significant improvement in the $H_0$ measurement despite an increased \gls{gw} catalog when adding in galaxy catalog information~\cite{LIGOScientific:2021aug, LIGOScientific:2025jau,LIGOScientific:2025yae, LIGOScientific:2025hdt}. 

Inferring $H_0$ from \gls{gw} observations is challenging since our estimates for the source properties depend on the theoretical waveform model used for inference; our models are not perfect descriptions of the full predictions of general relativity, and systematic errors due to mis-modelling known physical effects, often referred to as waveform systematics, can cause discrepancies in \emph{e.g.} the inferred luminosity distance and detector-frame component masses. 
For example, when analysing GW231123\_135430, a short duration \gls{gw} signal from the first part of the fourth \gls{gw} observing run~\cite{LIGOScientific:2025rsn}, the inferred luminosity distance varied from 0.7 -- 4.1~Gpc depending on the model used; each model inferred a large distance uncertainty: $\sim 2\, \mathrm{Gpc}$. However, typically the difference between models is much smaller. Refs.~\cite{Kunert:2024xqb,Dhani:2025xgt} explored the impact of waveform systematics on our ability to infer $H_0$ by injecting and recovering different models that include the same underlying physics. They showed that for a population of \glspl{bns}, the expected biases from different modelling frameworks -- effective-one-body~\cite{Pan:2013rra, Babak:2016tgq, Bohe:2016gbl, Pompili:2023tna,Cotesta:2018fcv, Cotesta:2020qhw,Ossokine:2020kjp, Ramos-Buades:2023ehm,Gamboa:2024hli}, phenomenological~\cite{Husa:2015iqa, Khan:2015jqa,Hannam:2013oca, Khan:2018fmp, Pratten:2020fqn,Estelles:2020osj,Estelles:2021gvs, Hamilton:2021pkf,London:2017bcn, Garcia-Quiros:2020qpx,Estelles:2020twz,Khan:2019kot, Pratten:2020ceb, Thompson:2023ase, Colleoni:2024knd} and post-Newtonian~\cite{Blanchet:1995ez,Blanchet:2004ek,Mishra:2016whh} -- are below statistical uncertainties of current detectors (using the bright siren method), while for \glspl{bbh} it is possible to obtain unreliable measurements for $H_0$ in future \gls{gw} observing runs.

Another source of systematic errors is the exclusion of known general relativistic phenomena when inferring the source properties from \glspl{gw}; typically, simplified models are initially developed and improved upon over subsequent years by incorporating additional physical effects. Two examples are spin-induced orbital precession -- the general relativistic phenomenon where spins misaligned with the orbital angular momentum of the binary causes the orbital plane to precess around the total angular momentum~\citep{Apostolatos:1994mx} -- and higher-order multipole moments~\citep{Goldberg:1966uu}.
A major benefit of using models that include spin-precession and higher-order multipoles is that they break degeneracies in the parameter space, particularly the distance -- inclination degeneracy~\citep{Usman:2018imj,CalderonBustillo:2020kcg}. Refs.~\cite{Feeney:2020kxk,Vitale:2018wlg} explored the impact of missing physics on our ability to infer $H_0$ for a population of neutron star black hole binaries with electromagnetic counterparts. Ref.~\cite{Feeney:2020kxk} showed that by including spin-precession in their models, a $\sim 40\%$ improvement in their estimates for $H_0$ can be obtained. Ref.~\cite{Vitale:2018wlg} demonstrated that more precise distance estimates can be obtained for precessing neutron star black hole binaries compared to non-spinning binary neutron star mergers. By assuming that the inferred redshift and peculiar velocity of the host galaxy were perfectly known, this leads to a reduced $H_{0}$ uncertainty.

In this work, we investigate how missing physics impacts our ability to infer $H_0$ for a population of \glspl{bbh} when using the mass spectrum method. By analysing both real and simulated \gls{gw} data, we show there is no significant advantage in performing inference with models that include spin-precession and higher-order multipole moments when estimating $H_0$; other statistical errors dominate the uncertainty. We verify that we only see minor differences in the inferred $H_0$ when simulating a highly spinning, non-astrophysically motivated population with preferences for asymmetric binary masses, where spin-precession and higher-order multipole moments are expected to contribute significantly to the \gls{gw} morphology. Our conclusions are based on both current and future sensitivities of the Advanced LIGO~\cite{LIGOScientific:2014pky} and Virgo~\cite{det4-VIRGO:2014yos} \gls{gw} detectors. 

This paper is structured as follows: in Sec.~\ref{sec:methods} we review the different waveform models used in this work, as well as single-event and hierarchical Bayesian analyses. In Sec.~\ref{sec:gwtc_reanalysis} we analyse publicly available \gls{gw} strain data with waveform models that include different underlying physics and investigate the impact on estimating $H_0$. In Sec.~\ref{sec:simulation} we consider how our conclusions change for a population of \glspl{bbh} with large spin magnitudes and preferentially asymmetric component masses; a region of the parameter space where spin-precession and higher-order multipoles are expected to significantly impact single-event Bayesian analyses. Finally, in Sec.~\ref{sec:conclusions} we conclude with a discussion of our results and their impact for future \gls{gw} analyses.

\section{Methods} \label{sec:methods}

\subsection{Waveform models} \label{sec:waveforms}

\begin{figure*}
    \centering
    \includegraphics[width=0.49\textwidth]{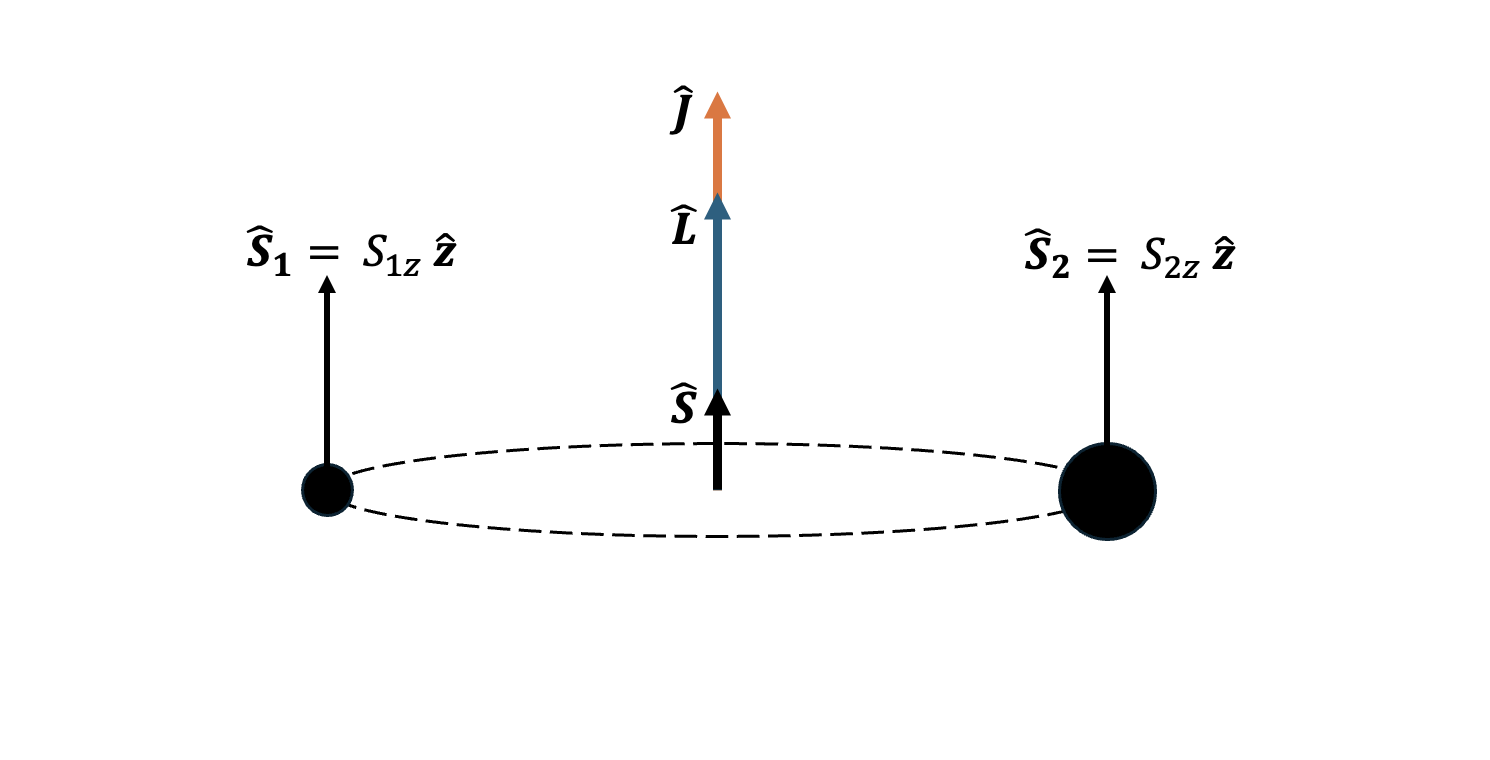}
    \includegraphics[width=0.5\textwidth]{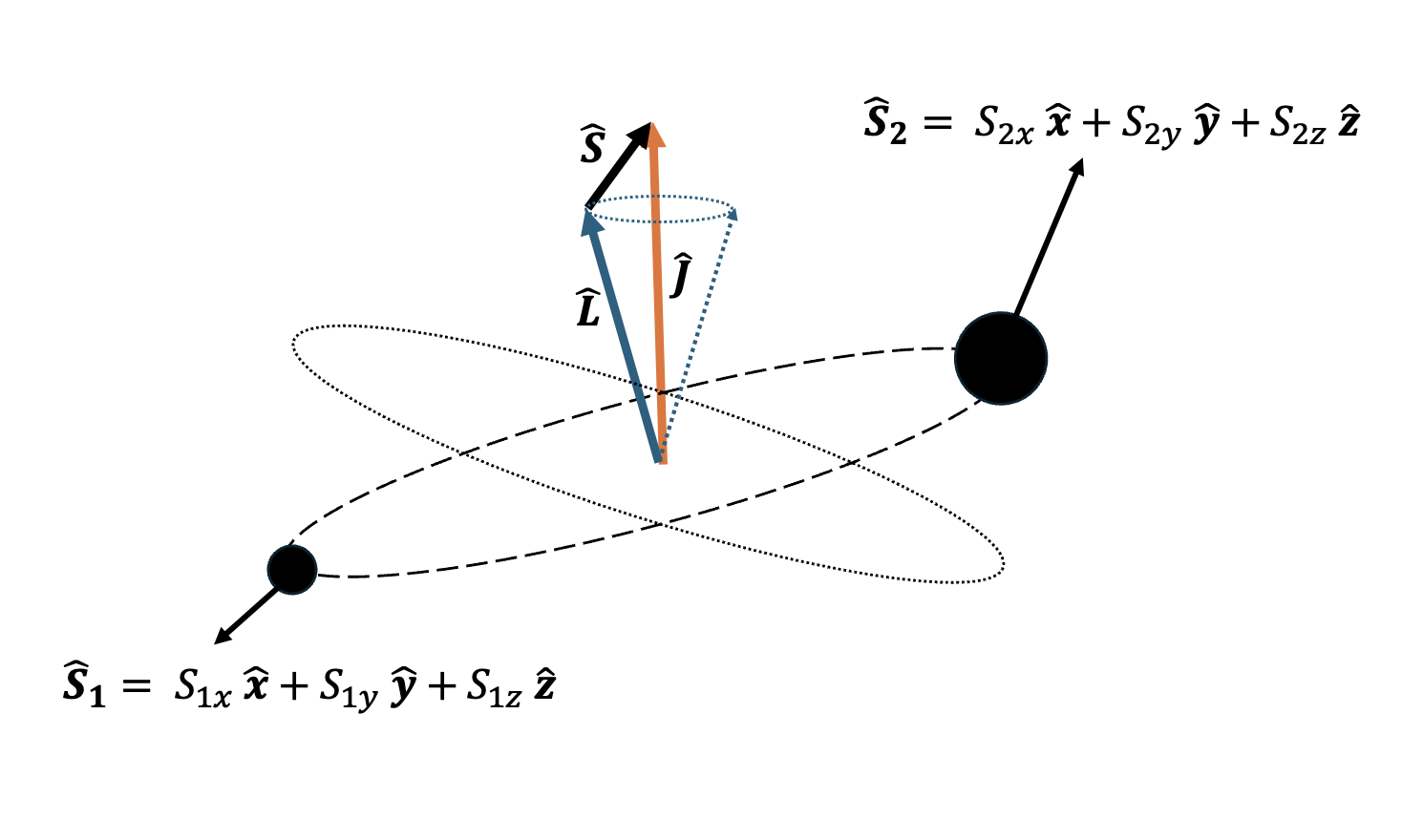}
    \caption{Illustration of a binary with \emph{Left}: spin angular momentum $\hat{\mathbf{S}} = \hat{\mathbf{S}}_{1} + \hat{\mathbf{S}}_{2}$ aligned with the orbital angular momentum $\hat{\mathbf{L}}$, and \emph{Right}: $\hat{\mathbf{S}}$ misaligned with $\hat{\mathbf{L}}$. When $\hat{\mathbf{S}}$ is misaligned with $\hat{\mathbf{L}}$ the binary undergoes the general relativistic phenomenon of spin-induced orbital precession: both $\hat{\mathbf{S}}$ and $\hat{\mathbf{L}}$ precess around the direction of the total angular momentum $\hat{\mathbf{J}} = \hat{\mathbf{L}} + \hat{\mathbf{S}}$ leaving visible amplitude modulations in the emitted \gls{gw}~\cite{Apostolatos:1994mx}. The dashed black line shows an instantaneous snapshot of the orbital plane. The dotted black line in the \emph{Right} panel indicates the path of $\hat{\mathbf{L}}$ as it precesses around $\hat{\mathbf{J}}$, and the dotted black line shows a snapshot of the orbital plane when $\hat{\mathbf{L}}$ has partially rotated around $\hat{\mathbf{J}}$. We choose a coordinate system such that $\hat{\mathbf{L}}$ is aligned with the $z$-axis.
    }
    \label{fig:binary_configurations}
\end{figure*}

Theoretical waveform models are used throughout \gls{gw} astronomy; parameterised models are vital for the detection of \glspl{gw} from matched filtering~\cite{Allen:2005fk, Cannon:2011vi, Babak:2012zx, Cannon:2012zt, DalCanton:2014hxh, Adams:2015ulm, Usman:2015kfa, Cannon:2015gha, Messick:2016aqy,Venumadhav:2019tad, Sachdev:2019vvd, Chu:2020pjv, Davies:2020tsx, Aubin:2020goo} and for inferring the source properties through parameter estimation~\cite{Veitch:2014wba,Ashton:2018jfp,LIGOScientific:2025yae}. To date, the \gls{lvk} Collaboration primarily uses signal models from the {\texttt{Phenom}}~\cite{Husa:2015iqa, Khan:2015jqa, London:2017bcn, Hannam:2013oca, Khan:2018fmp, Khan:2019kot,Pratten:2020fqn, Garcia-Quiros:2020qpx, Pratten:2020ceb,Estelles:2020osj, Estelles:2020twz, Estelles:2021gvs,Hamilton:2021pkf,Thompson:2023ase,Colleoni:2024knd}, {\texttt{SEOBNR}}~\cite{Bohe:2016gbl, Cotesta:2018fcv, Cotesta:2020qhw, Ossokine:2020kjp,Babak:2016tgq,Pan:2013rra,Pompili:2023tna,Ramos-Buades:2023ehm,Gamboa:2024hli} and {\texttt{NRSurogate}}~\cite{Varma:2019csw, Varma:2018mmi} waveform families. While the {\texttt{Phenom}} and {\texttt{SEOBNR}} models differ in their underlying assumptions, both are constructed from a combination of analytic and semianalytic approximations, as well as numerical relativity calculations. On the other hand, {\texttt{NRSurrogate}} models are solely constructed from numerical relativity simulations\footnote{Hybridised {\texttt{NRSurrogate}} models combine post-Newtonian and {\texttt{SEOBNR}} waveforms for the early inspiral with numerical relativity simulations for late inspiral, merger and ringdown. Hybridised {\texttt{NRSurrogate}} models are currently only available for aligned-spin systems~\cite{Varma:2018mmi}.}.

Developing waveform models is a computationally intensive and time-consuming process.
While systematic biases are expected for models that are incomplete descriptions of general relativity~\cite[see e.g.][]{CalderonBustillo:2015lrt,Ramos-Buades:2023ehm,Thompson:2023ase,Yelikar:2024wzm,MacUilliam:2024oif,Dhani:2024jja,Akcay:2025rve}, there is merit in using \gls{gw} models that exclude certain physical processes. 
For example, less accurate models are typically computationally cheaper to evaluate (while producing comparable results in some cases), and they can help quantify evidence for a physical process within the Bayesian framework~\cite[e.g.][]{Hoy:2022tst}. Less accurate models may also be required due to (computational) limitations in the underlying data-analysis method~\cite[e.g.][]{Harry:2016ijz,McIsaac:2023ijd}.

Recently, efforts have focused on the inclusion of subdominant multipole moments~\cite{Goldberg:1966uu} and spins misaligned with the orbital angular momentum~\cite{Apostolatos:1994mx}. For binaries with spins aligned with the orbital angular momentum, the majority of radiation is emitted in the dominant $(\ell, |m|) = (2, 2)$ quadrupole. However, additional power may also be radiated in subdominant multipole moments, including the $(\ell, |m|) = (3, 3)$ and $(4, 4)$ multipoles. When spins are misaligned with the orbital angular momentum $\mathbf{L}$, both the orbital angular momentum and spin angular momenta $\mathbf{S} = \mathbf{S}_{1} + \mathbf{S}_{2}$ precess around the direction of the total angular momentum $\mathbf{J}$, leaving visible amplitude modulations in the emitted \gls{gw}, see Fig.~\ref{fig:binary_configurations}.

Both the {\texttt{Phenom}} and {\texttt{SEOBNR}} waveform families have produced state-of-the-art models for the dominant quadrupole, spin-precession and/or higher-order multipole moments. Although each waveform family has its advantages and disadvantages, we use the ``generation X'' set of {\texttt{Phenom}} waveform models in this work: the frequency-domain {\texttt{IMRPhenomX}} models~\cite{Pratten:2020fqn}. We use,

\begin{itemize}
\item {\texttt{IMRPhenomXAS}}~\cite{Pratten:2020fqn} (XAS): XAS is limited to spins aligned with the orbital angular momentum and the dominant quadrupole,
\item {\texttt{IMRPhenomXP}}~\cite{Pratten:2020ceb} (XP): XP includes spins misaligned with the orbital angular momentum. XP uses XAS as the base model (in the co-precessing frame~\cite{Schmidt:2010it}) and applies a rotation to produce the precessing waveform in the detector frame,
\item {\texttt{IMRPhenomXHM}}~\cite{Garcia-Quiros:2020qpx} (XHM): XHM is limited to spins aligned with the orbital angular momentum and includes a subset of higher-order multipole moments with $\ell < 5$,
\item {\texttt{IMRPhenomXPHM}}~\cite{Pratten:2020ceb} (XPHM): XPHM is an amalgamation of the XP and XHM models. For this case, XHM is used as the base model (in the co-precessing frame) and a similar rotation is applied to the produce the precessing, higher-order multipole waveform in the detector frame.
\end{itemize}

The \gls{lvk} Collaboration, and others, primarily use models that include spin-precession and higher-order multipoles for analyses~\citep{Nitz:2021zwj,LIGOScientific:2021usb,LIGOScientific:2021djp,Olsen:2022pin,Mehta:2023zlk}; in the past, the \gls{lvk} Collaboration has analysed all \gls{bbh} candidates with at least XPHM~\cite{LIGOScientific:2021usb,LIGOScientific:2021djp,LIGOScientific:2025slb}. This is natural since it more accurately describes numerical solutions to general relativity~\cite{Pratten:2020ceb}. Also, the underlying astrophysical population of \glspl{bbh} remains unknown, and hence the waveform model should remain agnostic, \emph{i.e.} we should not assume black hole spin alignment or the (non-) observability of higher-order multipole moments when inferring their properties. However, with a greater understanding of the astrophysical formation mechanism of binary black holes, other approximants may be sufficient for certain signals. For example, in the isolated channel where binary black holes form from isolated stellar progenitors, spins are preferentially aligned with the orbital angular momentum~\citep{Kalogera:1999tq,Mandel:2009nx}.
In this case, XAS or XHM may be sufficient to reliably infer the source properties.

\subsection{Single-event parameter estimation}
\label{sec:single_event_pe}

The source properties, described by the multi-dimensional vector $\params = \paramsset$, are inferred from an observed \gls{gw} through Bayesian inference~\cite{Veitch:2014wba, Ashton:2018jfp}. For a quasi-circular binary black hole, $\params$ is a 15-dimensional vector including two dimensions for the component masses, three for each component spin angular momentum, two for the sky location, two for the luminosity distance and inclination angle of the source, and three for the geocentric merger time, phase, and polarization of the binary. 
In Bayesian inference, Bayes' theorem is exploited to calculate the \emph{posterior probability distribution}, defined as the probability of $\params$ conditional on the observed \gls{gw} data, $d$, and parameterised waveform model, $\model$. The posterior distribution is calculated through,

\begin{equation}
    p(\params | d, \model) = \frac{\mathcal{L}(d | \params, \model)\, \Pi(\params | \model)}{\mathcal{Z}(d, \model)},
\end{equation}
where $\mathcal{L}(d | \params, \model)$ is the probability of the data given the vector $\params$ and model $\model$, otherwise known as the likelihood, $\Pi(\params | \model)$ is the probability of the vector $\params$ given the model $\model$, otherwise known as the prior, and $\mathcal{Z}(d, \model)$ is $\int \mathcal{L}(d | \params, \model)\, \Pi(\params | \model)\, d\params$, otherwise known as the evidence. 
It is often not possible to analytically calculate the posterior distribution. For this case, stochastic sampling can be employed~\cite{metropolis1949monte,Skilling2004,Skilling:2006gxv}, although see \cite[e.g.][]{Pankow:2015cra,Lange:2018pyp,Delaunoy:2020zcu,Green:2020hst,Chua:2019wwt,Green:2020dnx,Dax:2021tsq,Gabbard:2019rde,Tiwari:2023mzf,Fairhurst:2023idl} for alternative methods.

\begin{figure*}
    \centering
    \includegraphics[width=0.98\textwidth]{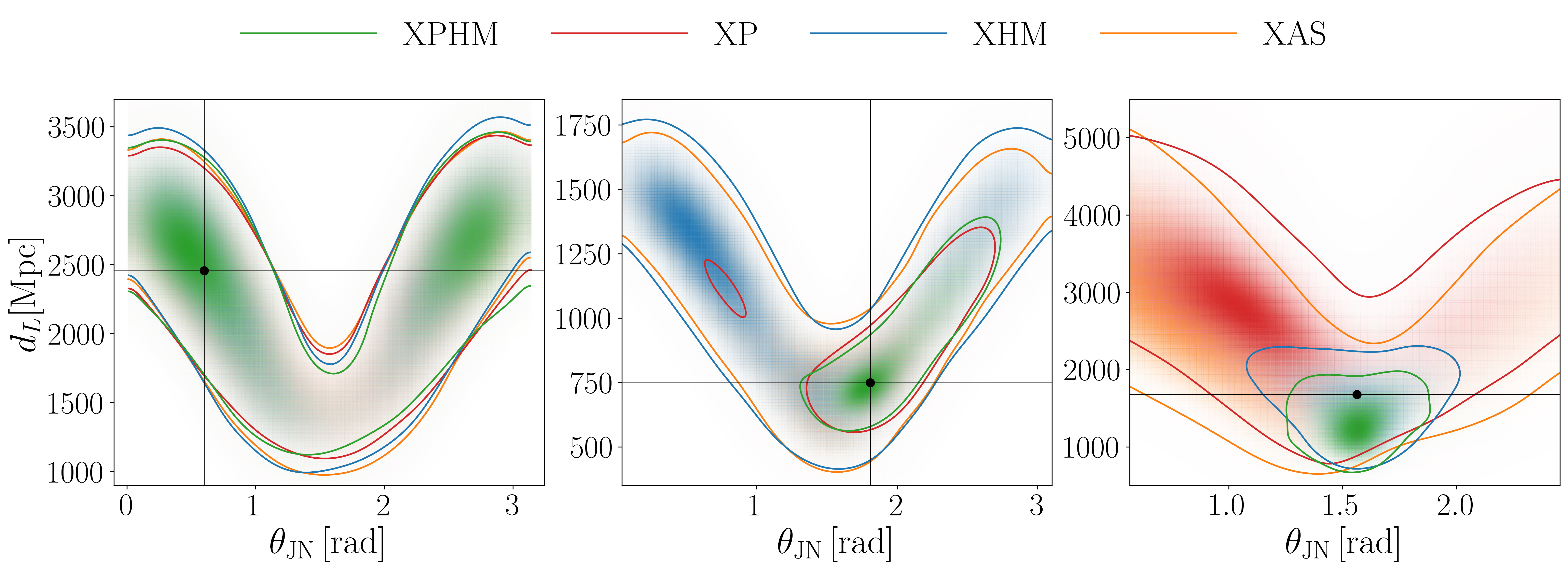}
    \includegraphics[width=0.98\textwidth]{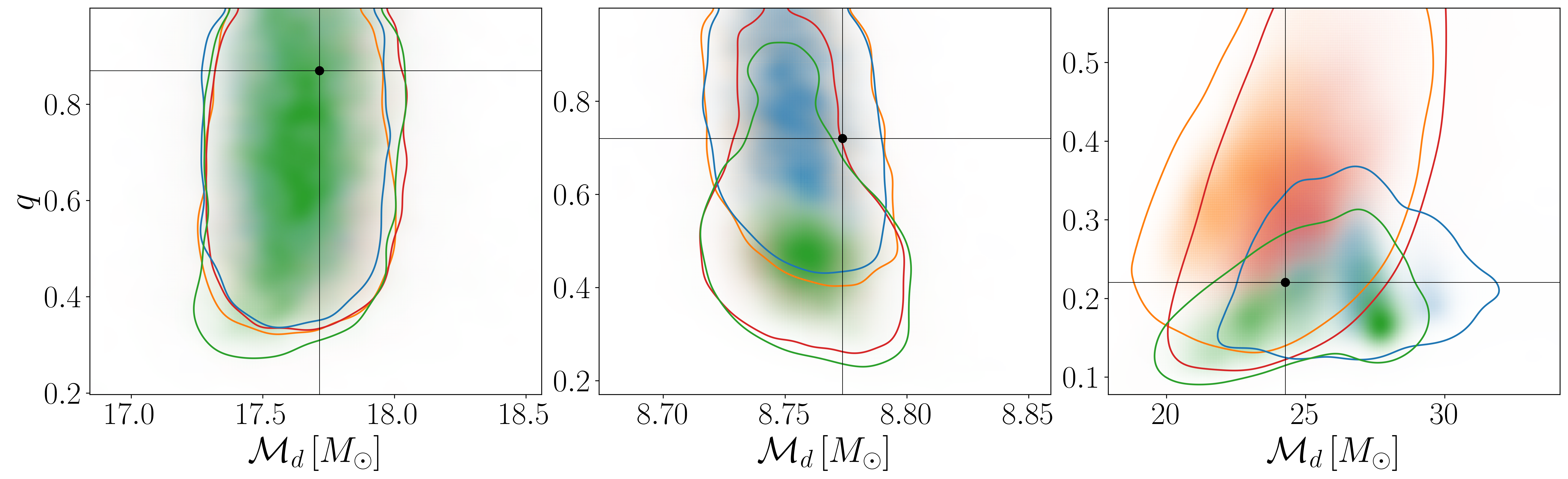}
    \caption{Two-dimensional marginalized posterior distributions for the inferred \emph{Top}: luminosity distance, $d_{\mathrm{L}}$, and inclination angle, $\theta_{\mathrm{JN}}$ and \emph{Bottom}: detector-frame chirp mass $\mathcal{M}_{d}$, and mass ratio $q$ for three simulated \gls{gw} signals injected into a three-detector network operating at their design sensitivities for the fourth GW observing run~\citep{O4PSD}. All simulated signals were produced with XPHM. Contours show the 90\% credible interval, and different colours represent posterior distributions obtained with different models, as indicated in the legend. The black cross hairs show the injected value. The \emph{Left} panel shows an analysis of an injection with no detectable precession and higher-order multipole moments, the \emph{Middle} panel shows an analysis of an injection with detectable precession, and the \emph{Right} panel shows an analysis of an injection with detectable higher-order multipole moments. Since these injections were analysed in idealised Gaussian noise, the peak of the posterior will not always lie at the true value, even when the model used for the injection and recovery is the same.
    }
    \label{fig:iota_distance_degeneracy}
\end{figure*}

Numerous packages are available to perform Bayesian inference on \gls{gw} signals~\cite{Veitch:2014wba,Ashton:2018jfp,Romero-Shaw:2020owr,Biwer:2018osg}. In this work, we take advantage of {\texttt{bilby}}~\cite{Ashton:2018jfp,Romero-Shaw:2020owr} and employ the {\texttt{Dynesty}} stochastic sampler~\cite{Speagle:2019ivv}. 
We consistently use 1000 live points for single-event parameter estimation. 

Performing Bayesian inference on \gls{gw} signals is challenging due to strong degeneracies in the parameter space~\citep[see e.g.][]{Baird:2012cu,Usman:2018imj}. These degeneracies can be broken by an observation of precession and/or subdominant multipole moments in the \gls{gw} signal, enabling more accurate estimates for the source properties~\citep[see e.g.][]{Kalaghatgi:2019log,Green:2020ptm,Mills:2020thr,Fairhurst:2023idl}. 
In Fig.~\ref{fig:iota_distance_degeneracy} we see that for signals with no observable precession and higher-order multipoles, the inferred posterior distribution does not change with the waveform model; although XPHM more accurately describes the true binary physics, for all models employed in this work we obtain the expected distance -- inclination degeneracy~\citep{Usman:2018imj,CalderonBustillo:2020kcg} and consistent estimates for the chirp mass and mass ratio of the binary. When the signal has detectable precession or higher-order multipoles, models that neglect this phenomenon show the previous degeneracy, but models that include it provide a more accurate inference for the \gls{gw} distance and inclination angle. For binaries with observable precession and higher-order multipoles, we additionally infer more support for comparable component mass ratios when spin-precession and higher-order multipoles are neglected. Fig.~\ref{fig:iota_distance_degeneracy} highlights that when precession and/or subdominant multipole moments are observable in the \gls{gw} signal, certain models can provide significantly tighter constraints on the inferred luminosity distance and mass ratio of the source. In principle, reduced uncertainties of the luminosity distance should imply tighter constraints on $H_0$.

\subsection{Cosmological parameter inference -- the mass spectrum method}
\label{subsec: hyperparameter inference}

Hierarchical Bayesian inference is used to estimate the posterior distribution of the hyperparameters $\hyper$ (including the cosmological parameters), given the observation of a GW catalog $\setdgw$. Through Bayes' theorem, the posterior $p(\hyper|\setdgw)$ can be computed through:

\begin{equation}\label{eq:bayes_hyper}
    p(\hyper|\setdgw, \model) = \frac{p(\setdgw|\hyper, \model)\,p(\hyper)}{p(\setdgw| \model)},
\end{equation}
where $p(\setdgw|\hyper, \model)$ is the hierarchical likelihood, $p(\hyper)$ is the hyperparameter prior and $p(\setdgw|\model)$ is the evidence. Note that we have explicitly retained the waveform model $\model$ dependence since we vary the model in this work.

Assuming all GW signals are of astrophysical origin and all noise realizations are independent, the hierarchical likelihood can be computed through\footnote{This expression marginalizes over the total rate of GW events.} \cite{Mandel:2018mve, Vitale:2020aaz, Mastrogiovanni:2021wsd}
\begin{equation}
    \label{eq: hierarchical likelihood}
    p(\setdgw|\hyper, \model)
    =
   \prod_{\nu=1}^{\nobs}
    \frac{\int\dd\params\,p(\datagwsub{\nu}|\params, \model)\,p(\params|\hyper)}{\int\dd\params\,\pdet\,p(\params|\hyper)}\,,
\end{equation}
where $\nobs$ is the number of observed events in the catalog, $\datagwsub{\nu}$ denotes the $\nu$th event in the catalog, and $\pdet$ is the detection probability of an event with source parameters $\params$.\footnote{The detection probability also depends on the waveform model used, but this difference is expected to be small. Since we are interested in the impact of the waveform model through data, not on the selection effects, we neglect this here and always use XPHM to quantify selection effects. }
The detection probability depends on how the catalog is constructed, for example the signal-to-noise-ratio (SNR) threshold that is applied to the catalog \cite{Essick:2023toz}, or other selection criteria that indicates the astrophysical origin of the events. 
Finally, the model for the parameters given the hyperparameters $\hyper$ is described by $p(\params|\hyper)$. 
$p(\params|\hyper)$ specifies the distribution of all GW source parameters, including the mass and spin model. 
A crucial part of the mass spectrum method is the choice of mass distribution. 
Later, we will analyze the data with the commonly-used \textsc{PowerLaw+Peak} distribution~\cite{LIGOScientific:2020kqk} for the primary component mass, which consists of a power law with a superimposed Gaussian distribution, the most preferred mass distribution from the third gravitational wave catalog (GWTC-3) data \cite{LIGOScientific:2021aug}.\footnote{This source-frame mass model has now been superseded by more complex mass models, including the \textsc{Multi Peak} model that has two Gaussian components and the \textsc{FullPop-4.0} model~\cite{Fishbach:2020ryj,Farah:2021qom,Mali:2024wpq} that encompasses the full mass spectrum of \glspl{cbc}. Both exhibit higher Bayes factors with the latest available data \citep{LIGOScientific:2025jau}. } 
Throughout, we assume the spins to be uniformly distributed over the sphere with uniform spin magnitude distribution, and the mass and spin models to be independent of redshift. 
The integral appearing in the numerator of Eq.~\eqref{eq: hierarchical likelihood} is usually approximated in the literature with Monte-Carlo sums over posterior samples of the individual GW events, for details see e.g.~\cite{Mastrogiovanni:2023emh}. In this work, we follow the same approach.

The detection probability -- $\pdet$ in the denominator of Eq.~\ref{eq: hierarchical likelihood} -- must be accurately modelled to avoid biased inference. The detection probability is required as current template-based search pipelines adopted by the \gls{lvk} Collaboration~\citep{Cannon:2011vi,
Privitera:2013xza,Messick:2016aqy,Hanna:2019ezx,
Sachdev:2019vvd,Usman:2015kfa,
Nitz:2017svb,Nitz:2018rgo,Adams:2015ulm,Chu:2020pjv,
Liu:2012vw,Guo:2018tzs} are not equally likely to observe all \gls{gw} signals; they are more likely to detect massive black hole binaries with spins aligned with the orbital angular momentum due to the increase in the sensitive volume. 
In fact, low \gls{snr} binaries with significant evidence for precession and higher-order multipole moments may remain undetected due to limitations in search algorithms\footnote{Current template-based search pipelines adopted by the \gls{lvk} restrict attention to spins aligned with the orbital angular momentum and neglect the presence of higher-order multipole moments (although see Refs.~\cite{McIsaac:2023ijd,Schmidt:2024jbp,Harry:2017weg,Chandra:2022ixv,Wadekar:2023gea,Klimenko:2008fu, Lynch:2015yin, Skliris:2020qax,Chandra:2020ccy})}. 
The detection probability can be estimated by simulating signals, injecting them into real GW data and identifying the fraction identified from the search pipelines. 
This assesses the performance of template-based search pipelines across the parameter space. 
To simulate signals, we are forced to make a reference choice of hyperparameters, $\hyper_{\rm ref}$, that determine how the injections are distributed in $\params$ space. 
If the injections cover the explored binary parameter space sufficiently, the final hyperparameter posterior is independent of the reference choice $\hyper_{\rm ref}$. 

For a total number of $\ninjtot$ of injected signals, and a probability of $p(\params|\hyper_{\rm ref})$ of being drawn in a reference population/cosmology, the denominator of Eq.~\eqref{eq: hierarchical likelihood} can be computed via a Monte-Carlo sum:
\begin{equation} \label{eq:likelihood_denominator}
    \int\dd\params\,\pdet\,p(\params|\hyper)
     \approx
     \frac{1}{\ninjtot}\sum_{\nu=1}^{\ninjsel}
     \frac{p(\params_{\nu}|\hyper)}{p(\params_{\nu}|\hyper_{\rm ref})}
     \,,    
\end{equation}
where the $\nu$ index runs over the entire recovered injection set (i.e.~the signals that pass the selection criterion), with $\ninjsel$ the number of recovered signals.

We note that the Monte-Carlo sum over posterior samples (in the numerator of Eq.~\ref{eq: hierarchical likelihood}), and the Monte-Carlo sum in the denominator of Eq.~\ref{eq: hierarchical likelihood}, are subject to fundamental statistical uncertainties. The expected variance in our estimate of $\log p(\setdgw|\hyper, \model)$ scales as~\cite{Essick:2022ojx, Talbot:2023pex, Heinzel:2025ogf}

\begin{equation} \label{eq:monte_carlo_error}
    \sigma^{2} \approx \sum_{\nu=1}^{\nobs} \sigma_{i}^{2} + \nobs^{2}\, \sigma_{\mathrm{det}}^{2} \,,
\end{equation}
where $\sigma^{2}_{i}$ is the variance from the individual posterior samples, and $\sigma_{\mathrm{det}}^{2}$ is the variance of Eq.~\ref{eq:likelihood_denominator}. As the variance from Eq.~\ref{eq:likelihood_denominator} scales quadratically with the catalog size, it is possible that we may not be able to reliably perform inference in the future with a large catalog of observed events; see~\cite{Wolfe:2025yxu} which suggests analysing fewer but higher \gls{snr} events as a solution.

\section{Results}

\subsection{Re-analysis of the \gls{bbh} mergers of GWTC-3}
\label{sec:gwtc_reanalysis}

To assess the impact of precession and higher-order multipole moments on the measurement of $H_0$, we re-analyze the \gls{bbh} mergers detected up to the third observing run\footnote{The \gls{lvk} Collaboration announced an additional 128 \gls{gw} candidates from the first part of the fourth observing run (O4a)~\citep{LIGOScientific:2025slb}. This enabled the \gls{lvk} to provide updated constraints on the inferred Hubble constant~\citep{LIGOScientific:2025jau}. Due to computational limitations, we did not expand our study to include these additional signals. However, we expect to obtain similar conclusions since comparable $H_0$ distributions were obtained when including and excluding data from O4a, see Fig.11 in~\citep{LIGOScientific:2025jau}.} \cite{LIGOScientific:2021djp} of the Advanced LIGO~\citep{LIGOScientific:2014pky} and Advanced Virgo~\citep{det4-VIRGO:2014yos} \gls{gw} detectors.
We use the same selection cuts as the original LVK cosmological inference work \cite{LIGOScientific:2021aug}, namely an SNR threshold of 11 and an Inverse False Alarm Rate (IFAR) threshold of 4~yr. This results in 42 \gls{bbh} events out of a total of $\sim 90$ observations.

The \gls{lvk} Collaboration released posterior samples obtained with XPHM for all 42 \gls{bbh} events in the GWTC-3 data release~\citep{LIGOScientific:2021djp}. To assess the impact of precession and higher-order multipoles we additionally require posterior samples obtained with XP and XHM. We obtained these posterior samples from~\cite{Hoy:2024wkc}. In this work, the authors re-analysed each \gls{gw} signal with XHM and XP, ensuring the same settings as the original \gls{lvk} analyses. As a result of using the posterior samples from~\cite{Hoy:2024wkc}, we did not re-analyse the \gls{bbh} mergers in GWTC-3 with XAS, although we will use XAS later in Sec.~\ref{sec:simulation}. However, given the lack of evidence for precession and higher-order multipole moments in the majority of events~\cite{Hoy:2024wkc}, the inferred $H_0$ from XAS will likely remain comparable to the other waveform models.

Given that we are analysing real GW candidates, the detection probability is directly related to the performance of the search pipelines adopted by the \gls{lvk} Collaboration. Owing to the significant computational cost for estimating the detection probability, we use publicly available results from the \gls{lvk} Collaboration~\cite{LIGOScientific:2021aug,ligo_scientific_collaboration_and_virgo_2021_5546676}. This dataset recovers $N=138,000$ (from a total of $2\times 10^8$ injected) signals, following the choices made by the LVK in~\cite{LIGOScientific:2021aug}.
To sample from the hierarchical posterior, we use the publicly available \textsc{icarogw} software package \cite{Mastrogiovanni:2023emh, Mastrogiovanni:2023zbw}. 
We use the \textsc{dynesty} sampler with 900 live points, and priors as defined in Appendix~\ref{app: priors}. 

\begin{figure}
    \centering
    \includegraphics[width=0.6\textwidth]{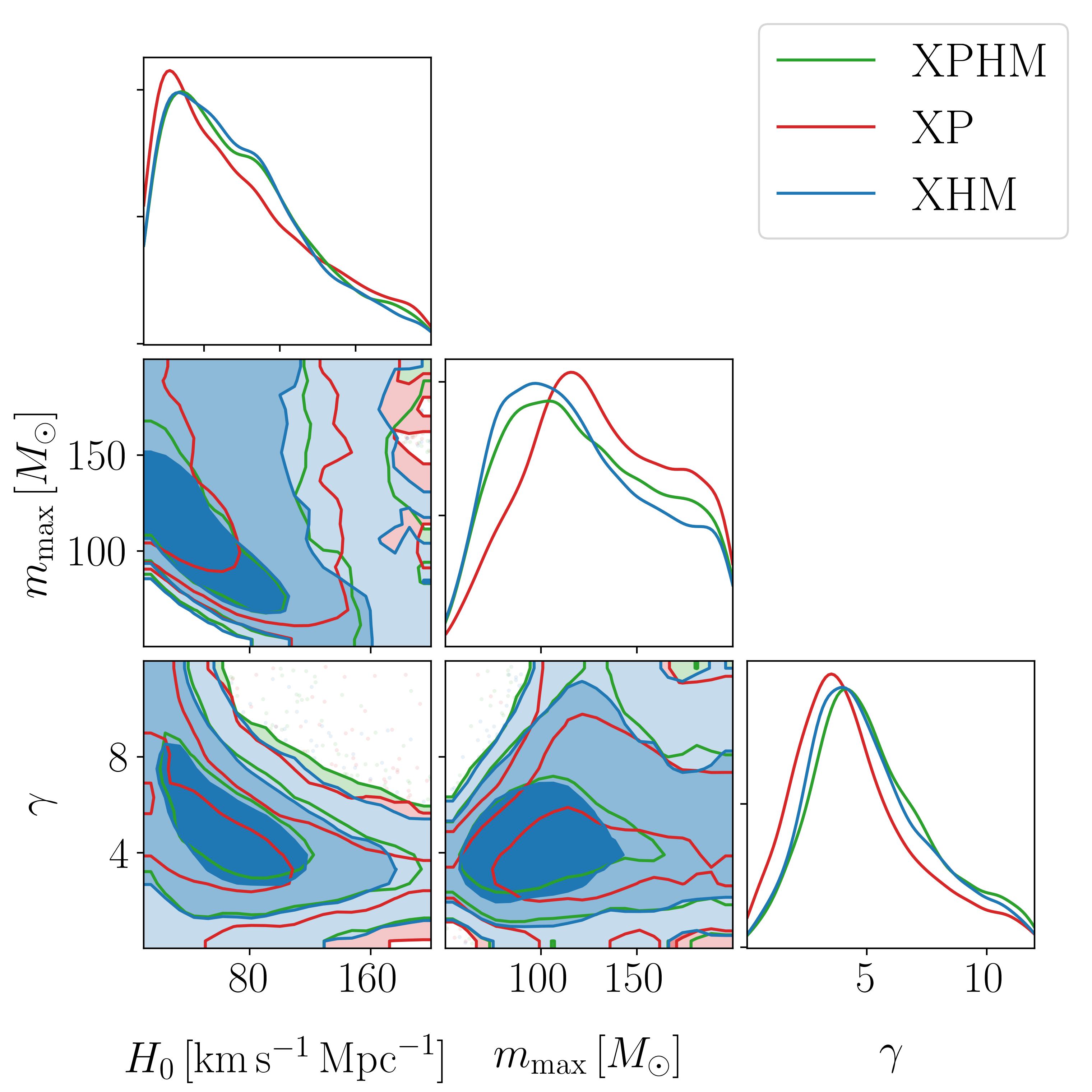}
    \caption{A subset of the cosmological and population parameters obtained when analyzing the GWTC-3 catalog of \glspl{bbh}. 
    The posteriors are colored according to the waveform assumed for the single-event PE. 
    The hyperparameters were selected since they are either of cosmological interest ($H_0$), correlate strongly with $H_0$ ($\gamma$, $\mmax$) or are most affected by the choice of waveform model ($\mmax$). $H_{0}$ is the Hubble constant, $\mmax$ is the maximum black hole mass and $\gamma$ is the inferred powerlaw of the redshift distribution, i.e.~$p(z) \propto (1 + z) ^ {\gamma - 1}  \tfrac{\mathrm{d}V_c}{\mathrm{d}z}$. See Appendix.~\ref{app: full results} for further details.
    }
    \label{fig: impact gwtc-3}
\end{figure}

The main result is shown in Fig.~\ref{fig: impact gwtc-3}. 
The inferred $H_0$ posterior shows negligible dependence on the physics included in the model; the addition of spin-induced orbital precession and higher-order multipoles produces minimal differences in the inferred cosmological parameters. We see small differences in the hyperparameters of the mass model used, for example the maximum black hole mass $m_{\mathrm{max}}$, but the 90\% credible intervals remain consistent. In general, we see that the lack of higher-order multipoles produces a marginally larger $m_{\mathrm{max}}$, which is correlated with the inferred power law of the mass ratio distribution, see Appendix~\ref{app:O3} -- in general, we see that models which exclude higher-order multipole moments prefer more symmetric mass ratio binaries when analysing the population. This is consistent with the single-event inference presented in Fig.~\ref{fig:iota_distance_degeneracy}: for binaries with significant evidence for higher-order multipole power, models that exclude this phenomenon tend to prefer more symmetric mass ratio binaries.

We quantify the statistical difference between marginalized $H_{0}$ posterior distributions by computing the Jensen-Shannon Divergence (JSD)~\cite{Lin:1991zzm}. The JSD varies between $0$ and $1\, \mathrm{bits}$, where $0\, \mathrm{bits}$ implies statistically identical, and $1\, \mathrm{bits}$ implies disjoint distributions. As described in~\cite{LIGOScientific:2018mvr}, a general rule of thumb is that a $\mathrm{JSD} < 5 \times 10^{-2}\, \mathrm{bits}$ implies that the distributions are in good agreement. We calculate JSDs of $4\times 10^{-3}\, \mathrm{bits}$ and $5\times 10^{-4}\, \mathrm{bits}$ when comparing XP to XPHM and XHM to XPHM respectively. 
Since this remains less than the naive $5\times 10^{-2}\, \mathrm{bits}$ criterion, and comparable to the expected JSD from re-performing the same analysis with a different random seed (JSD $\sim 10^{-3}\, \mathrm{bits}$)~\cite{Dax:2021tsq}, we conclude that this difference is not significant. 

\begin{figure}
    \centering
    \includegraphics[width=1.0\textwidth]{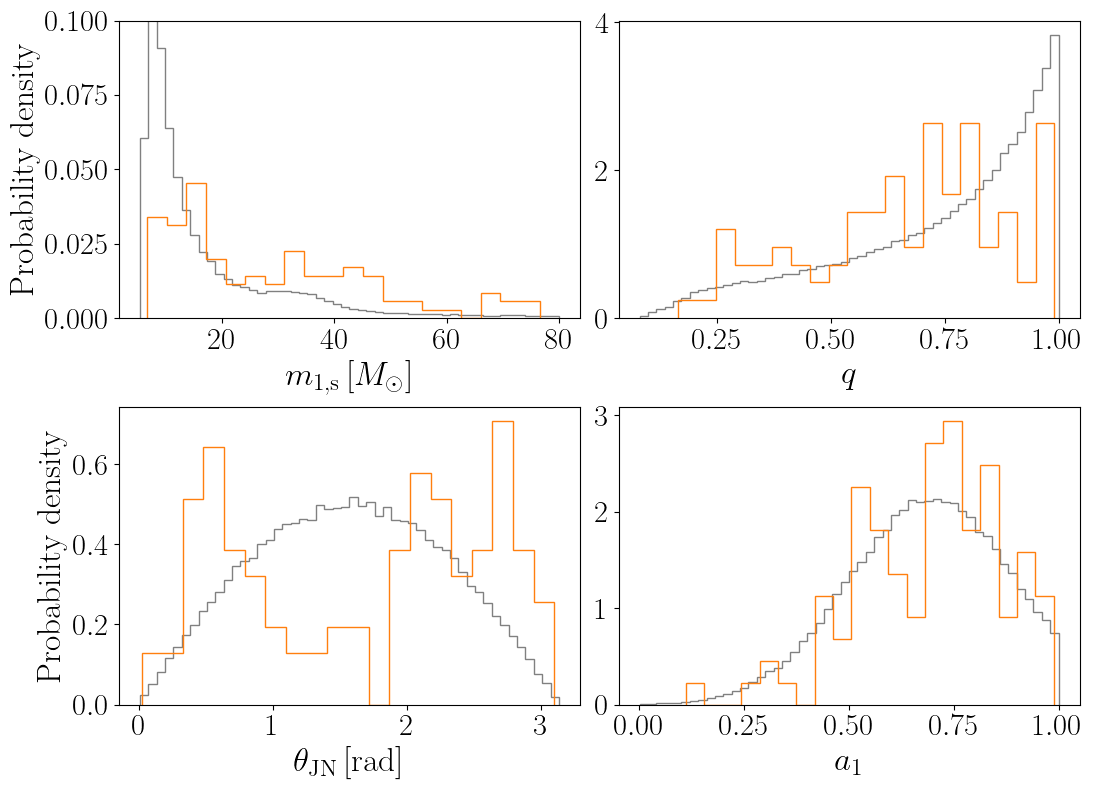}
    \caption{In grey we show the properties of 100,000 binaries drawn from a highly spinning, preferentially asymmetric component mass population of \glspl{bbh}. In orange we show the 80 randomly chosen injections that have SNR $> 12$, analysed in Sec.~\ref{sec:simulation}. The \emph{Top Left} panel shows the source frame primary mass, the \emph{Top Right} panel shows the mass ratio, the \emph{Bottom Left} panel shows the inclination angle of the binary and the \emph{Bottom Right} panel shows the primary spin magnitude.
    }
    \label{fig:highspin_population}
\end{figure}

\subsection{Simulated population} \label{sec:simulation}

Given that the inferred $H_0$ shows very little dependence on precession and higher-order multipole moments for real \gls{gw} signals, we further test whether this is a consequence of the observed population, and whether our conclusions remain with improved \gls{gw} detectors. We simulate a population of \glspl{bbh} with significant spin-precession and higher-order multipole power. We estimate that 8\% of the binaries drawn from the observed \gls{lvk} population in GWTC-3 are included in this simulated population; the remaining binaries are more symmetric with smaller spins. This simulated population can therefore be considered a ``worst-case'' scenario.

We achieve this simulated population by assuming the mass ratio follows a power law $q^{\beta_{q}}$ with index $\beta_{q} = -1.1$, and spin magnitudes follow a truncated Gaussian distribution with mean $a_{i} = 0.7$ and standard deviation $\sigma=0.2$. The other population parameters are inspired from the GWTC-3 population~\citep{KAGRA:2021duu}. Specifically, we assume that the primary mass follows a truncated power law with index $\alpha=-2.3$, between a minimum, $m_{\mathrm{min}} = 5\, M_{\odot}$, and maximum $m_{\mathrm{max}} = 80\, M_{\odot}$ mass. We also assume that there is a Gaussian component in the mass distribution at $\msone = 34\, M_{\odot}$, with relative weight $\lambda = 0.038$ and standard deviation $\sigma_{\rm g} = 5~M_{\odot}$. 
Finally, we model the redshift distribution as a power-law relation $(1 + z)^{\gamma}$ with $\gamma=2.0$ that modulates the usual uniform-in-comoving distance distribution, i.e.~$p(z)\propto \tfrac{\mathrm{d}V_c}{\mathrm{d}z}\tfrac{(1 + z)^2}{1 + z} $, where the $1 / (1 +z)$ accounts for the time dilation between detector and source-frame. 
Fig.~\ref{fig:highspin_population} shows the properties of $100,000$ binaries drawn from our reference population, and the subset of binaries after imposing the selection threshold (and therefore accounting for the detection probability).

Although the mass ratio distribution for the drawn binaries may be expected to peak away from unity due to the power-law distribution assumed, Fig.~\ref{fig:highspin_population} shows that the mass ratio histogram is skewed towards equal-mass binaries. 
This is a consequence of most binaries in the population having small mass; if the heavier mass is close to the minimum assumed black hole mass, this by definition produces $q\approx 1$ binaries. 
Note that after imposing the selection criterion the mass ratio distribution is significantly skewed away from unity. This is because heavier binaries have a higher detection probability (which are more likely to have lower $q$). 
As such, the interpretation of $\beta_{q} = -1.1$ is challenging from Fig.~\ref{fig:highspin_population} because of the chosen parametrization -- the minimum bound of $q$ depends on $\msone$, which makes the link between the marginal $q$ distribution (shown above) and the conditional $p(q|\msone)$ non-trivial. This is further discussed in Appendix~\ref{app:simulation}.

Rather than injecting signals into real GW data and performing a search to identify which are detected, we simplify the process by assuming signals with matched-filter \glspl{snr} greater than 12 are detected -- a common practice \cite{Agarwal:2024hld}. 
This greatly simplifies the calculation for the detection probability: we draw $\ninjtot$ simulated signals from our reference population, calculate the \gls{snr} of each \gls{gw} signal, and consider those with \gls{snr} $> 12$ as being observed. The detection probability for a simulated signal with parameters $\injparams$ is then a Heaviside step function: 

\begin{equation}
    \pdet = 
    \begin{cases}
        0 & \mathrm{if} \,\, \rho(\injparams) < 12 \\
        1 & \mathrm{if} \,\, \rho(\injparams) \geq 12
    \end{cases}
\end{equation}
where,

\begin{equation}
    \rho(\params_{\mathrm{inj}}) = \frac{(\model(\params_{\mathrm{inj}}) | d)}{|\model(\params_{\mathrm{inj}})|} \,,
\end{equation}
$(a | b)$ is the inner product of the Fourier-transformed \gls{gw} strain weighted by the detector sensitivity~\cite{Finn:1992wt,Owen:1995tm}, $|a| = \sqrt{(a | a)}$ and $\model(\params_{\mathrm{inj}})$ corresponds to the theoretical \gls{gw} signal with binary parameters $\params_{\mathrm{inj}}$ produced by the model $\model$.
In this work, we simulate each GW signal with XPHM, and consider a single realisation of idealised instrumental noise (otherwise known as gaussian noise) based on the \gls{psd} of the detector; the specific realisation differs for each injection. 
We assume a three detector network operating at a 75\% duty cycle per detector. Our assumed network comprised of two Advanced LIGO detectors at their design sensitivity for the fourth GW observing run~\citep{O4PSD}, and one Virgo detector at its sensitivity at GPS time: 1249852257.0 (14th August 2019). 

\begin{figure}
    \centering
    \includegraphics[width=0.6\textwidth]{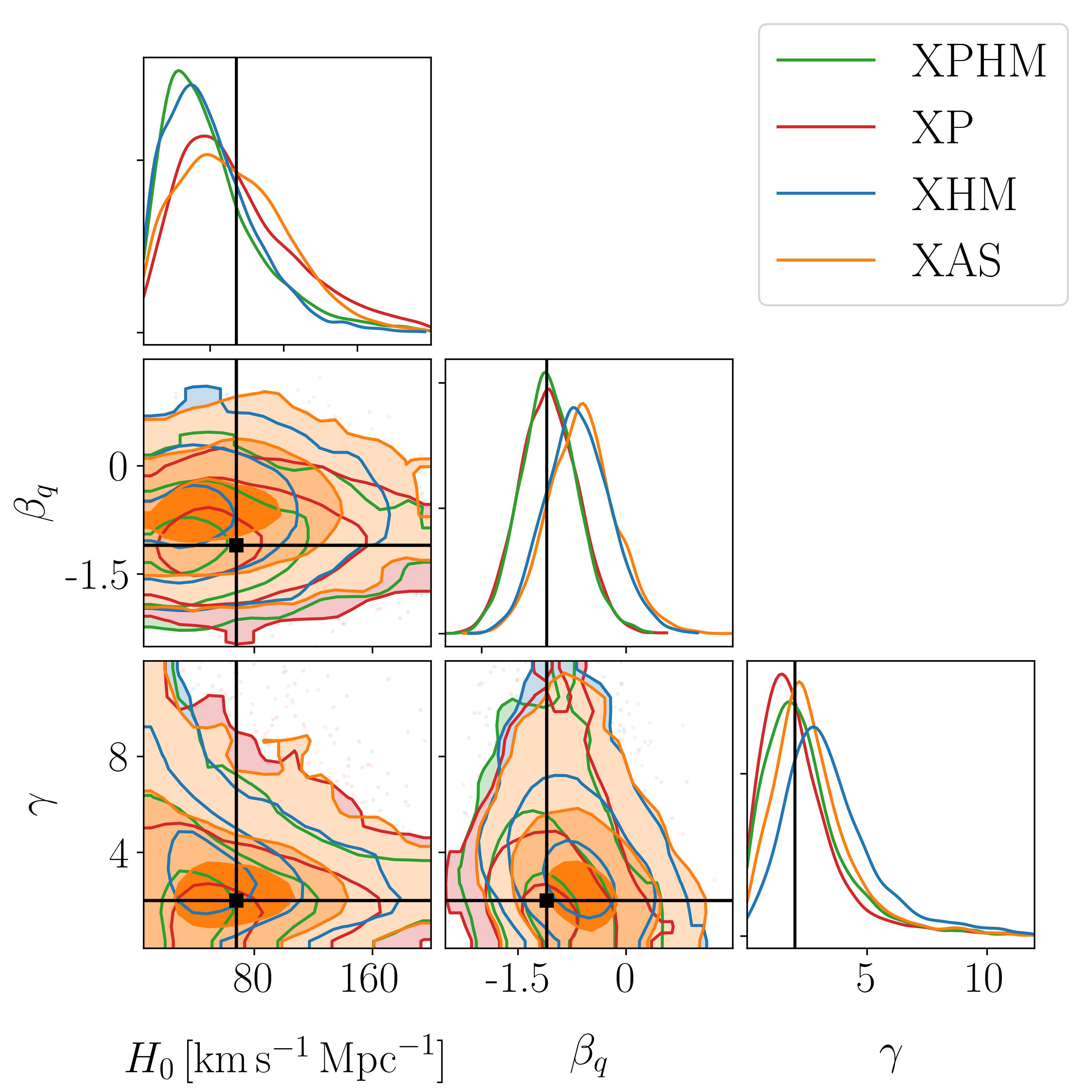}
    \caption{A subset of the cosmological and population parameters that are most affected by the choice of waveform model when analysing 80 \gls{gw} signals drawn from a fiducial population with high spins and preferentially asymmetric-mass binaries. The posteriors are colored according to the waveform assumed for the single-event PE and the black cross hairs show the true population values. $H_{0}$ is the Hubble constant, $\beta_{q}$ is the powerlaw of the mass ratio distribution and $\gamma$ is the powerlaw of the redshift distribution. See Appendix.~\ref{app: full results} for further details.
    }
    \label{fig: impact simulated}
\end{figure}

We produced a set of 80 detected binary mergers from the reference population, with Fig.~\ref{fig:highspin_population} showing their properties. Note, a population of 80 signals with SNR $> 12$ remains $\sim 2\times$ more than that observed from the first to third \gls{gw} observing runs~\citep{LIGOScientific:2021djp}, and more than all loud signals (SNR $> 12$) observed up until the first part of the fourth observing run~\citep{LIGOScientific:2025slb}. Since it has been shown that high \gls{snr} events contribute disproportionally to the resulting hyperparameter posterior \citep{Wolfe:2025yxu}, this population is representative of current and future event catalogs.

For each of the 80 detected \gls{gw} signals, we perform Bayesian inference following the details in Sec.~\ref{sec:single_event_pe}. 
Each signal is analysed with the models $\boldsymbol{\model} = \{\mathrm{XAS}, \mathrm{XP}, \mathrm{XHM}, \mathrm{XPHM}\}$.
From the corresponding single-event posterior samples we then compute the resulting hyperparameter posterior distributions via the cosmological inference pipeline \textsc{icarogw} \citep{Mastrogiovanni:2023emh} to determine $H_{0}$.
The final result is four different $H_0$ posterior distributions for each of the waveform models. 
All priors are detailed in Appendix~\ref{app: priors}. 

In general, we observe consistent conclusions as in Sec.~\ref{sec:gwtc_reanalysis}, \emph{i.e.} the inferred $H_0$ remains largely agnostic to the underlying physics in the model. In Fig.~\ref{fig: impact simulated} we see that all models recover the fiducial $H_0$ (used when generating the population) within the $1\sigma$ credible interval, and all 90\% credible intervals are consistent. For the $H_0$ inference we see a small correlation between models that include and exclude higher-order multipole moments: XPHM and XHM agree well, as do XAS and XP, but there are small differences between XPHM and XP.
This indicates that while neglecting precession has no impact, neglecting higher-order mode content is more important. This is supported by the JSDs; we infer statistical differences of $4\times 10^{-2}\, \mathrm{bits}$, $5\times 10^{-3}\, \mathrm{bits}$ and $4\times 10^{-2}\, \mathrm{bits}$ when comparing XP to XPHM, XHM to XPHM and XAS to XPHM respectively. This remains less than the naive $5\times 10^{-2}\, \mathrm{bits}$ rule of thumb, indicating that the $H_{0}$ distributions are in good agreement.

The inferred value of the mass ratio power-law exponent, $\beta_q$, follows a different trend to $H_0$. 
The value of $\beta_q$ is most strongly dependent on whether a model includes precession (rather than higher-order modes as previously for $H_0$).
This is understood from Fig.~\ref{fig:iota_distance_degeneracy} -- neglecting precession for a precessing signal leads to a tail to more symmetric mass ratios, $q\sim 1$. 
Therefore, the inferred $\beta_q$ distribution should be biased towards more equal-mass binaries since we assume a highly spinning population in our catalog; $\sim 60\%$ of our population has observable precession~\cite{Fairhurst:2019vut}.
Fig.~\ref{fig:iota_distance_degeneracy} seems to imply that neglecting higher-order multipole content should also bias the mass estimates, yet this is not seen in Fig.~\ref{fig: impact simulated}; we see that the $\beta_q$ posterior remains the same when including and excluding higher-order multipole power. 
It appears that even though the single-event posteriors can be biased when omitting higher-order modes, there must be a significant number of events with strong higher-order multipole power in the population to produce significant biases in the overall inferred mass ratio distribution; for instance, our population contained $25\%$ of binaries with observable higher-order multipole power~\cite{Mills:2020thr}. This implies that the inclusion of spin-precession may be required for inferring the mass distribution of black holes, while the impact of higher-order multipoles remains negligible. Note that this is also consistent with \citep{Singh:2023aqh}, which finds that higher-order mode content is not relevant for population studies for populations up to $\sim 700$ events. 
We stress that even though the waveform model that neglects precession infers a biased mass ratio distribution, this bias does not impact the measured $H_0$ value. 

It is expected that analyses with $\mathcal{O}(1000)$ binaries in the population should exhibit a stronger $H_0$ bias since the hyperparameter uncertainties shrink. Note, a population of this size may be subject to the Monte-Carlo error appearing in the population likelihood described in Eq.~\ref{eq:monte_carlo_error}. 
However, we highlight that our assumed population is astrophysically unlikely and significantly different from that inferred to date~\citep{LIGOScientific:2025pvj}. 
In particular, it contains highly spinning \glspl{bbh} that aggravate differences between models. 
Given that a more astrophysical population with little observable precession and higher-order multipole content is likely to contain these binaries as a subset, a more astrophysically-motivated population will likely exhibit a much smaller degree of $H_0$ bias.
To demonstrate this, we perform our analysis on a simulated population, now with astrophysically motivated parameters that predict a low-spin population, see Appendix~\ref{sec:population_low_spin} for details. 
We see that all differences between waveform models are now negligible and we obtain remarkably consistent results as in Sec.~\ref{sec:gwtc_reanalysis}.
The parameter with the largest scatter for the different waveform models is the power-law exponent governing the secondary mass, $\beta_q$ (similarly to before). Here, however, the bias on $H_0$ (and all other parameters) is negligible.

The average computational cost when analysing each of these 80 binaries with XAS, XP, XHM and XPHM was 30, 90, 55 and 180 CPU days respectively\footnote{All analyses were performed on AMD EPYC 2GHz nodes with 1 terabyte memory. To optimise throughput, the number of CPUs used for each analysis varied between 16 and 128.}. This implies that by simply using XAS for inference, comparable $H_0$ posteriors can be obtained in six times less total computational cost.

\section{Conclusions} \label{sec:conclusions}

In this work, we explore the impact of missing physics (higher-order multipoles and precession in \glspl{gw}) on our ability to infer $H_0$ with \gls{gw} observations via the mass spectrum method. 
Using full Bayesian inference for both the individual and hierarchical analyses, we show that even for a population with large, non-aligned spin (leading to spin-precession) that has a larger fraction of \glspl{bbh} with unequal mass ratios (leading to observable higher-order multipoles), there is no advantage in using models that include physics beyond the dominant quadrupole to perform cosmological inference. 
We note that this population is a conservative choice not motivated by current astrophysical data -- the observed \gls{lvk} GW population has mostly low spin magnitudes with sources being equal mass-ratio.
Statistical uncertainties in individual \gls{gw} analyses dominate over errors from missing physics. Using simpler, less accurate models to infer $H_0$ reduces the total computational cost by a factor of six.
Although evidence for precession and higher-order multipoles in the \gls{gw} helps break the well known distance-inclination degeneracy, enabling more accurate estimates for the luminosity distance of the source, we argue that simpler models still obtain roughly the correct distance, although with a larger uncertainty. 
As apparent in the results from this study, it seems the exact single-event posterior shape is not relevant when estimating $H_0$ from GW data via the mass spectrum method.
Note that this no longer applies when direct electromagnetic counterparts are simultaneously detected -- in this case accurate spin-precession models are important~\cite{Feeney:2020kxk}.

Our results demonstrated that neglecting higher-order multipole information is more important than neglecting precession when estimating $H_0$. Further work is needed to identify the exact number of observed GW events above which this becomes an important $H_0$ systematic. However, given that we start to see minor differences in the inferred $H_0$ for a population including $48/80$ binaries with significant spin-precession and $20/80$ events with significant higher-order multipole power, we predict that we will need to observe a catalog of $\sim 1500$ binaries before we observe similar systematics in the observed population: \cite{Hoy:2024wkc} demonstrated that a single event with significant evidence for precession will occur once
in every $\sim 50$ detections, and a binary with significant evidence for higher-order multipoles will
occur once in every $\sim 70$ observations according to current estimates for the black hole mass and spin distribution. This means that we will need to detect $\sim 1500$ binaries before we observe a comparable number of binaries with significant higher order multipole power.

As part of this work, we also found that we obtained comparable sky localizations between waveform models when analysing most of the simulated signals in our fiducial population. Since accurately inferring the sky location is crucial to narrow down possible host galaxies, it is possible that our key conclusion -- there is little advantage in using models that include physics beyond the dominant quadrupole to perform cosmological inference -- extends to the dark siren method, which combines \gls{gw} observations with galaxy catalogs. Although we highlight that this is not explicitly tested in this work.

This work has consequences for the future of \gls{gw} astronomy: as we observe significantly more signals with next generation \gls{gw} detectors~\cite{Punturo:2010zz, Reitze:2019iox}, we may not be able to consistently use the most accurate models for Bayesian inference as the model accuracy typically correlates with computational cost. Unless novel methods are developed for performing Bayesian inference on \gls{gw} data in $\mathcal{O}({\rm sec})$, such as machine learning algorithms~\cite{Green:2020dnx,Green:2020hst,Dax:2021tsq,Dax:2024mcn,Raymond:2024xzj} or others~\cite{Fairhurst:2023idl}, we may be forced to use simpler models that include less physics. Thankfully, in the near future, we will still be able to reliably infer $H_0$ with \gls{gw} observations via the mass spectrum method.

\section*{Acknowledgments}

We thank Tessa Baker, Ian Harry and Laura Nuttall for helpful discussions and comments. We are also grateful to Alexander Papadopoulos for comments during the LVK internal review. C.H. thanks the UKRI Future Leaders Fellowship for support through the grant MR/T01881X/1, and the University of Portsmouth for support through the Dennis Sciama Fellowship. K.L. is supported by ERC Starting Grant SHADE (grant no.\,StG 949572). Numerical computations were carried out on the SCIAMA High Performance Compute (HPC) cluster which is supported by the Institute of Cosmology and Gravitation (ICG) and the University of Portsmouth. 

This research has made use of data or software obtained from the Gravitational Wave Open Science Center (gwosc.org), a service of the LIGO Scientific Collaboration, the Virgo Collaboration, and KAGRA. This material is based upon work supported by NSF's LIGO Laboratory which is a major facility fully funded by the National Science Foundation, as well as the Science and Technology Facilities Council (STFC) of the United Kingdom, the Max-Planck-Society (MPS), and the State of Niedersachsen/Germany for support of the construction of Advanced LIGO and construction and operation of the GEO600 detector. Additional support for Advanced LIGO was provided by the Australian Research Council. Virgo is funded, through the European Gravitational Observatory (EGO), by the French Centre National de Recherche Scientifique (CNRS), the Italian Istituto Nazionale di Fisica Nucleare (INFN) and the Dutch Nikhef, with contributions by institutions from Belgium, Germany, Greece, Hungary, Ireland, Japan, Monaco, Poland, Portugal, Spain. KAGRA is supported by Ministry of Education, Culture, Sports, Science and Technology (MEXT), Japan Society for the Promotion of Science (JSPS) in Japan; National Research Foundation (NRF) and Ministry of Science and ICT (MSIT) in Korea; Academia Sinica (AS) and National Science and Technology Council (NSTC) in Taiwan. This material is based upon work supported by NSF's LIGO Laboratory which is a major facility fully funded by the National Science Foundation.

\appendix

\section{Priors}
\label{app: priors}

When performing hierarchical Bayesian inference on a catalog of \gls{gw} signals $\setdgw$, we must assume a prior distribution on the hyperparameters $\hyper$: $p(\hyper)$. In this work, we consistently assume uniform distributions, $\mathcal{U}$, with wide bounds. Namely, for $H_0$ we assume $\mathcal{U}(5, 200)\, \mathrm{km}\, s^{-1}\, \mathrm{Mpc}^{-1}$, for the power law of the redshift distribution we assume $\gamma: \mathcal{U}(0, 12)$, and for the high redshift tapper we assume $\kappa: \mathcal{U}(0, 6)$. In all cosmological analyses we assume a fixed matter density parameter consistent with the Planck 2015 results~\cite{Planck:2015fie}: $\Omega_{m} = 0.308$.

The mass spectrum method simultaneously infers cosmological, and mass and spin hyperparameters. In this work, we assume a \textsc{Powerlaw+Peak} model for the primary component mass, unless otherwise stated, which forms a truncated power law with a Gaussian component, see Appendix~\ref{app:simulation} for details. The power law index follows a prior distribution $\alpha: \mathcal{U}(1.5, 12)$ with minimum black hole mass $m_{\mathrm{min}}: \mathcal{U}(2, 10)\, M_{\odot}$, maximum black hole mass $m_{\mathrm{max}}: \mathcal{U}(50, 200)$, mean of the Gaussian component $\mu_{g}: \mathcal{U}(20, 50)\, M_{\odot}$, width of the Gaussian component $\sigma_{g}: \mathcal{U}(0.4, 12)$, and a range of mass tapering at the lower end $\delta_{m}: \mathcal{U}(0.05, 10)\, M_{\odot}$. The mass ratio distribution is assumed to be a power law with index following a prior distribution $\beta_{q}: \mathcal{U}(-4, 12)$. We finally assume that the component spin magnitudes are independently and identically drawn from a Beta distribution~\cite{Wysocki:2018mpo} with prior distributions $\mathcal{U}(1.1, 5)$ for the mean and variance.

For the single-event parameter estimation analyses performed in Sec.~\ref{sec:simulation}, we assume standard priors used by the \gls{lvk} Collaboration. Specifically, we assume a jointly uniform prior in the detector-frame component masses with chirp mass bounds between 2 and $200\, M_{\odot}$ and mass ratio bounds between 0.05 and 1. The spin vectors are assumed to be isotropic on the sphere and uniform in spin magnitude. This prior is also used for models that enforce spins to be aligned with the orbital angular momentum: XAS, XHM. We use an isotropic prior for the location of the source on the sky and the distance prior is
uniform in comoving volume and source frame time. The single-event parameter estimation analyses assume a flat $\Lambda\mathrm{CDM}$ cosmology.

\section{Full results}
\label{app: full results}

\subsection{O3 results}
\label{app:O3}

\begin{figure*}
   \centering
   \includegraphics[width=0.635\linewidth]{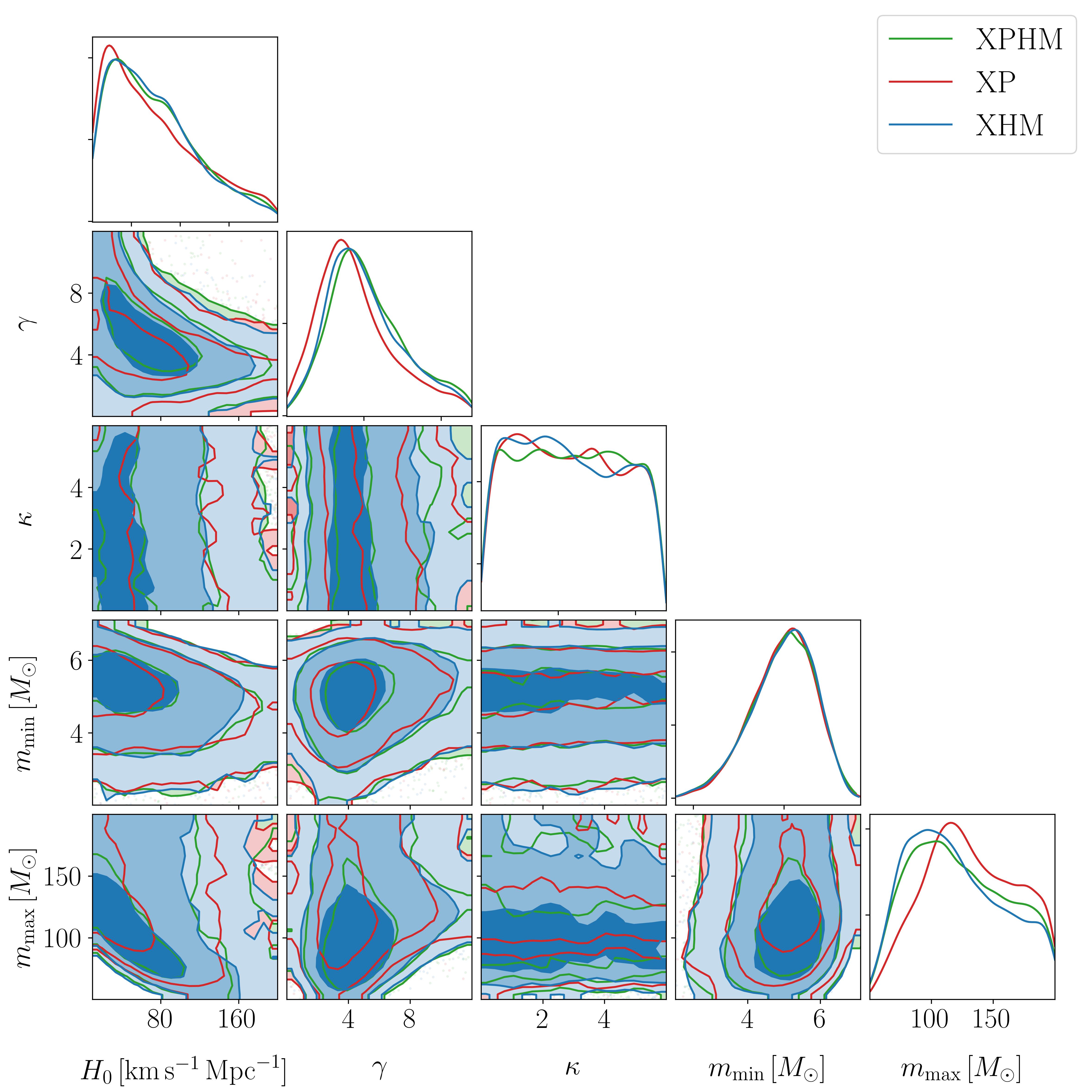}
   \includegraphics[width=0.635\linewidth]{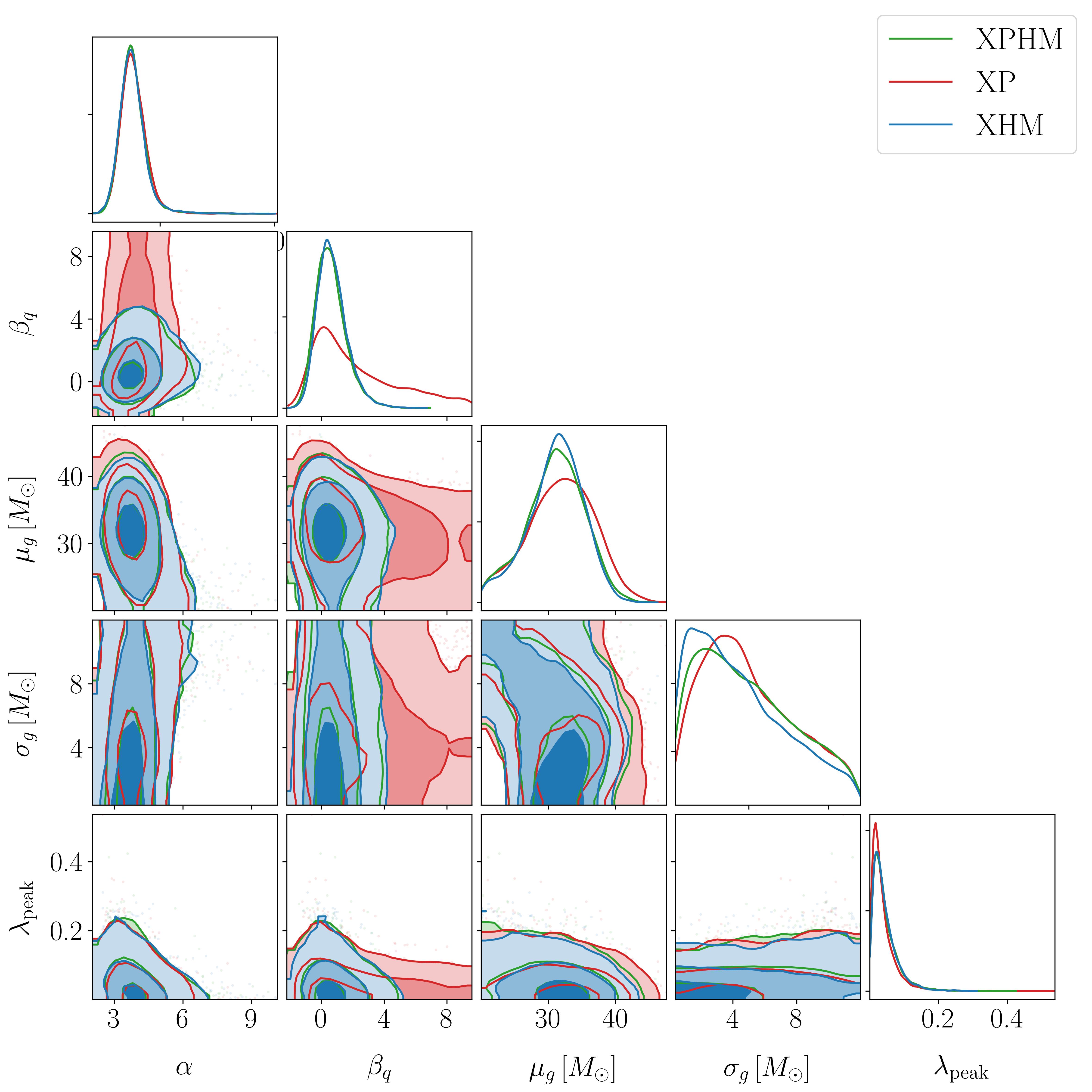}
   \caption{Most of the cosmological and population parameters analysing the GWTC-3 catalog of \glspl{bbh}. The posteriors are colored according to the waveform assumed for the single-event PE. In the \emph{Top} panel we show parameters that are either of cosmological interest or correlate strongly with $H_0$. In the \emph{Bottom} panel, we show parameters that are directly related to the mass model.}
   \label{fig: full corner O3}
\end{figure*}

In our re-analysis of \glspl{bbh} mergers detected up to the third observing run in Sec.~\ref{sec:gwtc_reanalysis}, we found negligible dependence on the choice of waveform model for the inferred $H_0$ posterior, see Fig.~\ref{fig: impact gwtc-3}. In Fig.~\ref{fig: full corner O3} we show the posterior distributions for the other cosmological and population parameters. We see that for all but one marginalized distribution, the posterior distributions agree well. We see small differences in the posterior distributions obtained for the mass ratio power law, $\beta_{q}$, with XP inferring a tail to larger $\beta_{q}$. This implies an overall preference for more symmetric mass ratio binaries in the population compared with models that include higher-order multipole effects: XPHM and XHM. Given that the median of the XP posterior remains consistent with the other models, and~\cite{Singh:2023aqh} demonstrated that neglecting higher-order multipole moments will not cause a significant bias in our inference of the underlying mass distribution for $\sim 700$ \gls{gw} signals, we suspect this tail is a consequence of the limited number of signals in GWTC-3, and more signals would reduce support for larger $\beta_{q}$. Indeed, as explained in Appendix~\ref{sec:population_low_spin} and shown in Fig.~\ref{fig: full corner O4 low spin}, the inferred $\beta_{q}$ remains consistent between XHM, XP and XPHM when the number of \gls{gw} signals in the catalog is doubled.

\subsection{Simulated O4 results (high spin)}
\label{app:simulation}

When simulating a population of \glspl{bbh} in Sec.~\ref{sec:simulation}, we assume a \textsc{Powerlaw+Peak} model for the primary component mass, and a power law for the mass ratio distribution. In {\textsc{icarogw}}, the mass model is factorised as~\cite{icarogw_tech}:

\begin{equation}
    p(\msone, \mstwo |\hyper) = p(\msone|\hyper)\, p(\mstwo| \msone, \hyper) S(\msone | \delta_{m}, m_{\mathrm{min}}) S(\mstwo | \delta_{m}, m_{\mathrm{min}}),
\end{equation}

where $S(a | \delta_{m}, m_{\mathrm{min}})$ is a window function defined as,

\begin{equation}
    S(a | \delta_{m}, m_{\mathrm{min}}) = 
        \begin{cases}
            0 & (a < m_{\mathrm{min}}) \\
            \left[f(a - m_{\mathrm{min}}, \delta_{m}) + 1\right]^{-1} & m_{\mathrm{min}} \leq a < m_{\mathrm{min}} + \delta_{m} \\
            1 & m \geq m_{\mathrm{min}} + \delta_{m} \\
        \end{cases}
\end{equation}

and,

\begin{align}
    [f(m', \delta_{m}) + 1]^{-1} & = \exp\left(\frac{\delta_{m}}{m'} + \frac{\delta_{m}}{m' - \delta_{m}}\right) \\
    p(\msone | \hyper) & = \frac{(1 - \lambda) \msone^{-\alpha}}{N} + \lambda \mathcal{G}(\msone | \mu_{g}, \sigma_{g}) \\
    p(\mstwo | \msone, \hyper) & = \frac{1}{N} \mstwo^{\beta_{q}}, \label{eq:powerlaw}
\end{align}
where $\mathcal{G}$ represents a Gaussian distribution, $\lambda$ is the mixing fraction and $N$ is a normalization factor.

\begin{figure}
    \centering
    \includegraphics[width=0.6\linewidth]{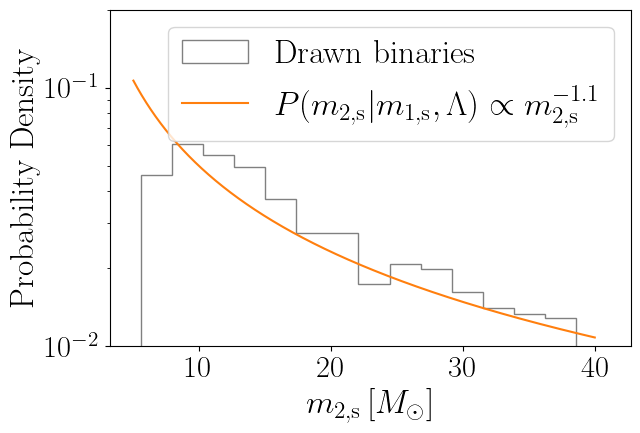}
    \caption{In grey we show the secondary masses for 100,000 binaries drawn from a highly spinning, preferentially asymmetric component mass population of \glspl{bbh}. In orange we show the expected conditional distribution given the assumed mass ratio model. 
    In this example, we fix $\msone=40~M_{\odot}$. 
    }
    \label{fig: conditional_m2_distribution}
\end{figure}

In Fig.~\ref{fig:highspin_population} we show the properties of 100,000 randomly chosen binaries from a reference population with significant spin-precession and higher-order multipole power. We see that the mass ratio distribution is skewed towards equal-mass binaries, despite a power-law distribution assumed with index $\beta_{q} = -1.1$. In Fig.~\ref{fig: conditional_m2_distribution} we show the conditional $p(\mstwo | \msone, \hyper)$ distribution. We see that the distribution for the secondary component mass follows the expected power law distribution from Eq.~\ref{eq:powerlaw}. The skew to more equal mass ratios is therefore a consequence of most binaries in the population having a low total mass, where the mass ratio is pushed towards unity due to the minimum black hole mass assumed for both components.

\begin{figure*}
   \centering
   \includegraphics[width=0.635\linewidth]{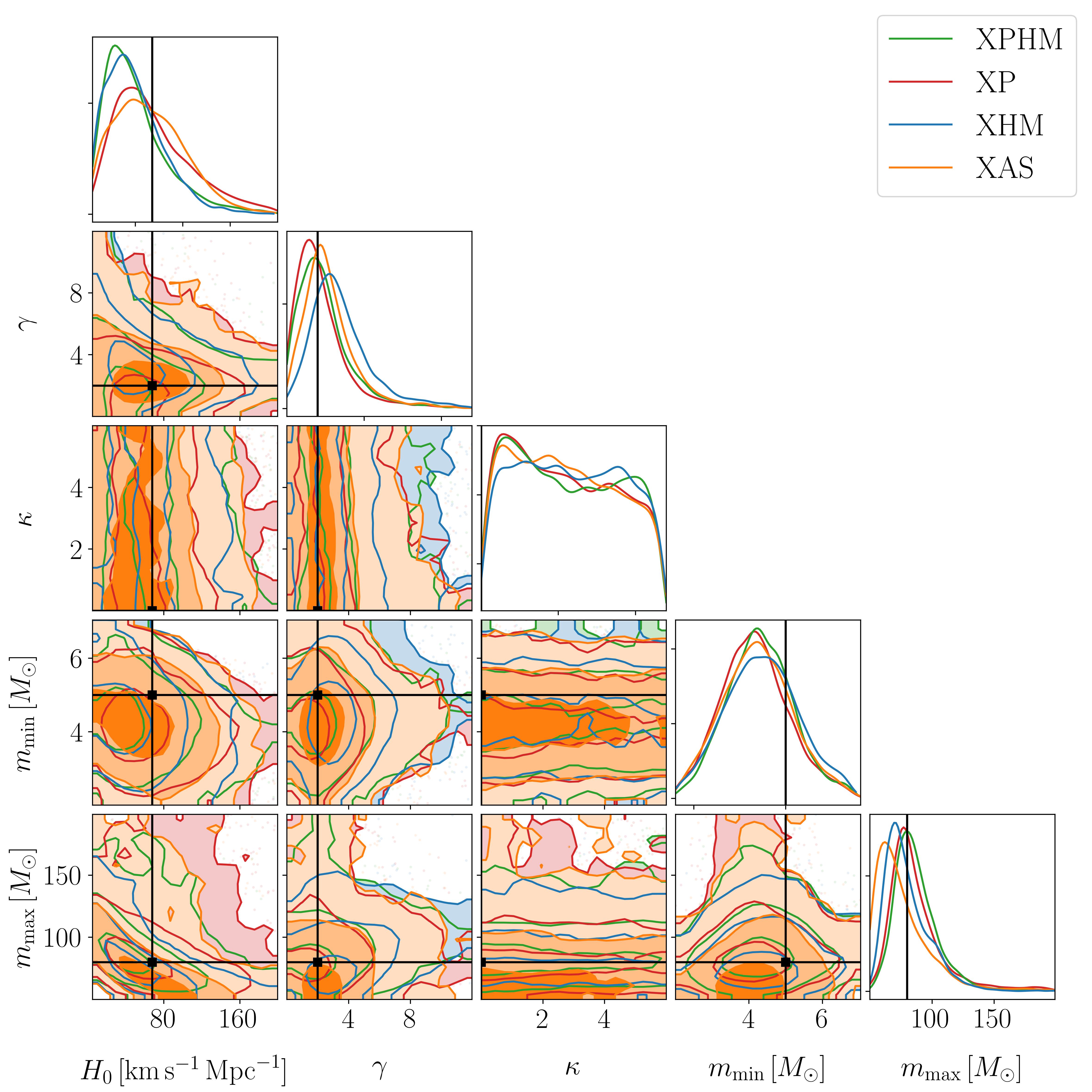}
   \includegraphics[width=0.635\linewidth]{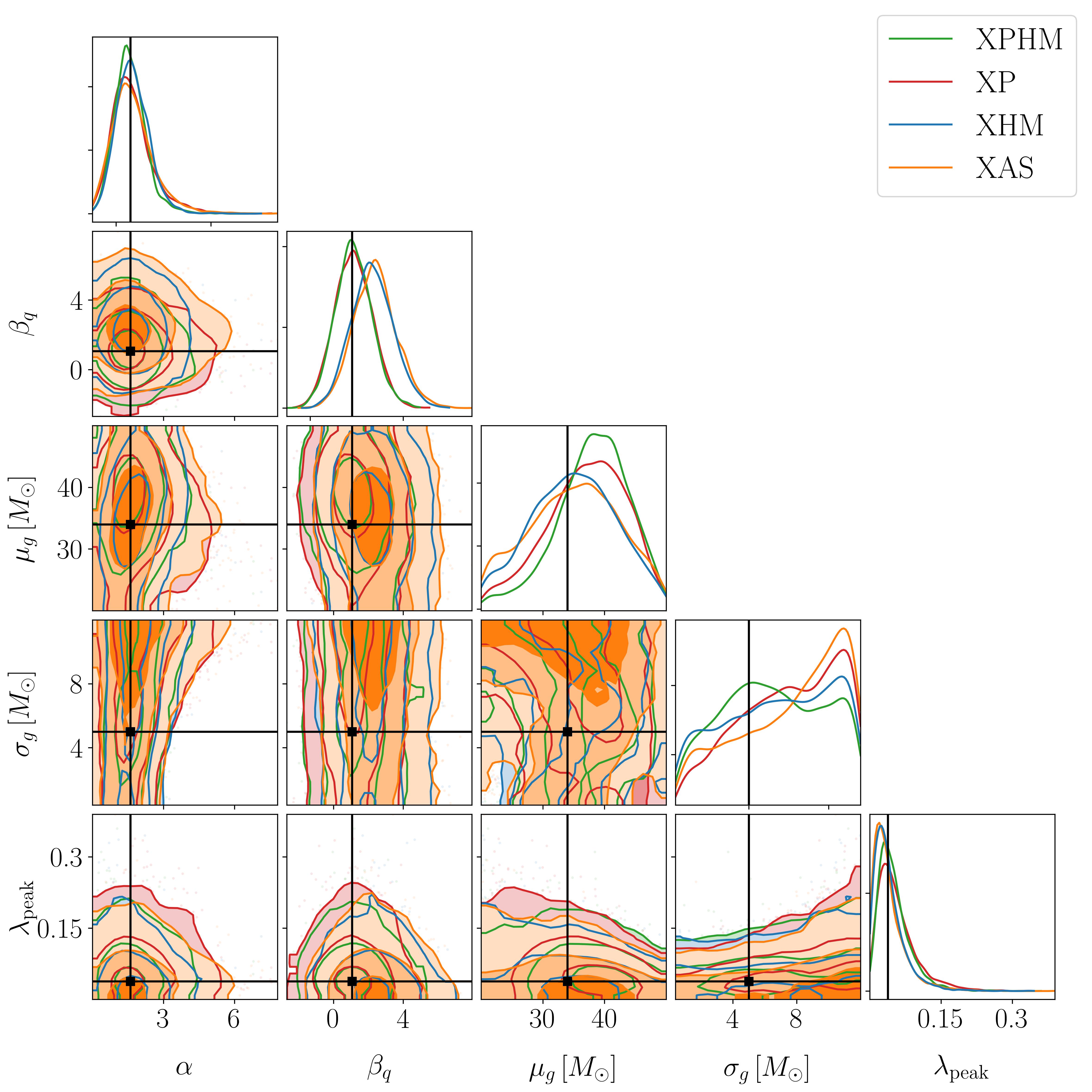}
   \caption{Similar to Fig.~\ref{fig: full corner O3} except we now show the cosmological and population parameters analysing the simulated high-spin population. The black vertical lines show the true population hyper parameters.
   }
   \label{fig: full corner O4 high spin}
\end{figure*}

Once 80 binaries are drawn from the reference population, and single-event parameter estimation performed on each signal, hierarchical inference is carried out to obtain posterior distributions for the cosmological parameters. In Fig.~\ref{fig: impact simulated} we show a subset of cosmological and population parameters that are most affected by the choice of waveform model. In Fig.~\ref{fig: full corner O4 high spin} we show posterior distributions for the other parameters. In general, we see good agreement between posterior distributions obtained with the different models. This further demonstrates that although we simulated a non-astrophysical population of \glspl{bbh} with significant evidence for precession and higher-order multipole power,  our ability to constrain cosmological parameters through the mass spectrum method remains larger agnostic to the underlying physics in the model.

\begin{figure*}
   \centering
   \includegraphics[width=0.95\linewidth]{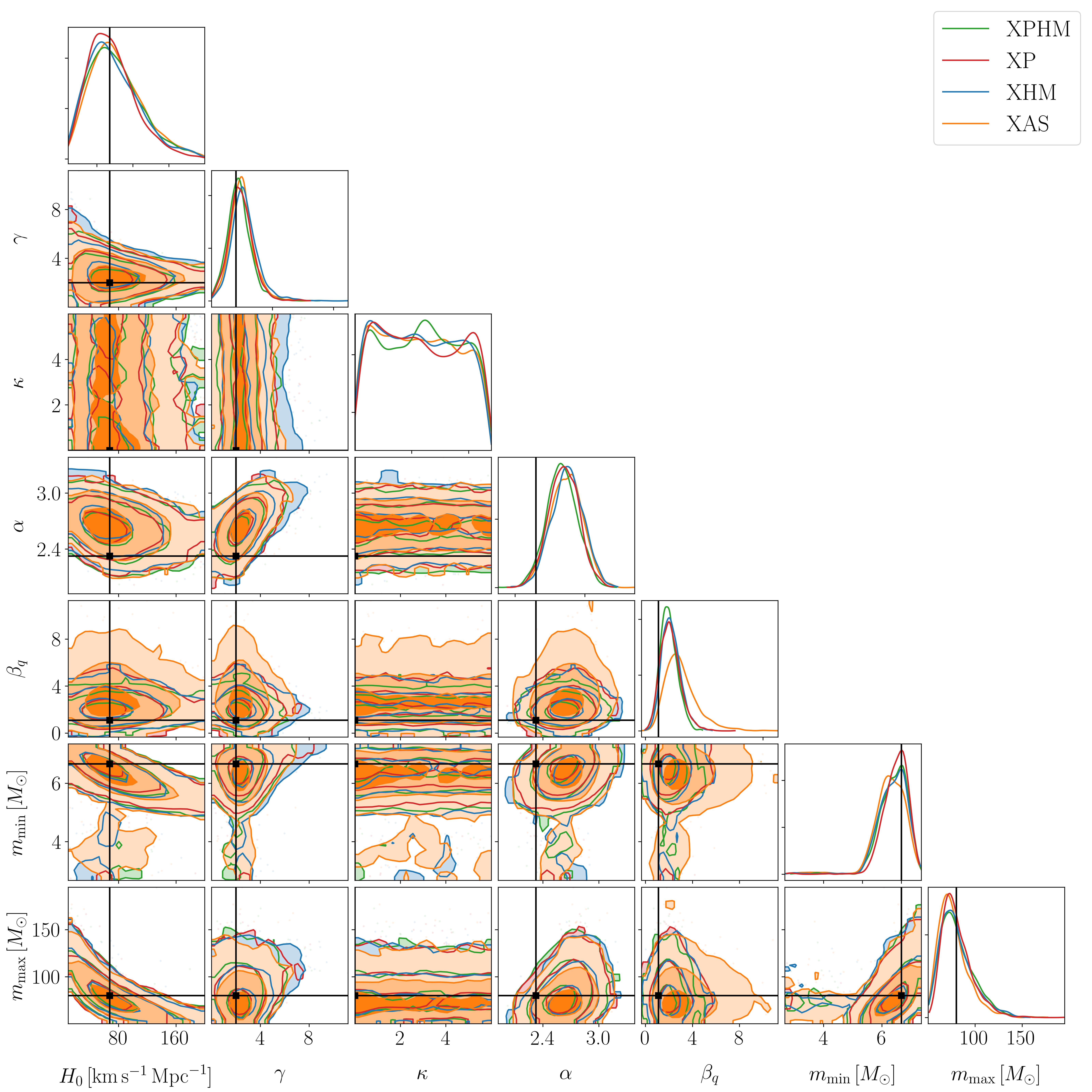}
   \caption{Similar to Fig.~\ref{fig: full corner O3} except we now show the cosmological and population parameters analysing the simulated low-spin population. The black vertical lines show the true population hyper parameters.}
   \label{fig: full corner O4 low spin}
\end{figure*}

\subsection{Simulated O4 results (low spin)} \label{sec:population_low_spin}

In addition to simulating a population of \glspl{bbh} with significant spin-precession and higher-order multipole power, as discussed in Sec.~\ref{sec:simulation} and Appendix~\ref{app:simulation}, we further consider an astrophysically motivated population of \glspl{bbh}. This population represents the signals observed by the \gls{lvk} Collaboration. Given that this population shows significantly lower evidence for precession and higher-order multipole power~\cite{Hoy:2024wkc}, we expect to observe negligible differences between inferences with XAS, XHM, XP and XPHM. We simulated a population where the mass ratio follows a power law $q^{\beta_{q}}$ with index $\beta_{q} = 1.1$, and a simple \textsc{Powerlaw} distribution for the primary component mass with index $\alpha = -2.3$ between a minimum, $m_{\mathrm{min}} = 5\, M_{\odot}$, and maximum $m_{\mathrm{max}} = 80\, M_{\odot}$. For this population, we assumed uniform spin magnitudes rather than a truncated Gaussian. Finally, we model the redshift distribution as a power-law relation $(1 + z)^{\gamma}$ with $\gamma=2.0$ that modulates the usual uniform-in-comoving distance distribution, i.e.~$P(z)\propto \tfrac{\mathrm{d}V_c}{\mathrm{d}z}\tfrac{(1 + z)^2}{1 + z} $, where the $1 / (1 +z)$ accounts for the time dilation between detector and source-frame. 
The other population parameters are inspired from the GWTC-3 population~\citep{KAGRA:2021duu}. As in Sec.~\ref{sec:simulation}, we randomly drew $\sim O(100)$ binaries from the population with $\mathrm{SNR} > 12$ and analysed each with the \textsc{dynesty} sampler via {\textsc{bilby}}. Population inference was then performed via {\textsc{icarogw}} using the priors described in Appendix~\ref{app: priors}.

As shown in Fig.~\ref{fig: full corner O4 low spin}, we see remarkable agreement between the posterior distributions obtained with the different models. We conclude that for an astrophysically motivated distribution of \glspl{bbh}, our ability to constrain $H_0$ through the mass spectrum method remains agnostic to spin-precession and higher-order multipoles. Unlike in Sec.~\ref{sec:simulation} and Fig.~\ref{fig: impact simulated} where we saw some small, non-significant differences in the posterior distribution for $H_0$, no disagreement is seen for this population. It seems likely that even if the number of binaries in our catalog significantly increases to $\sim O(1000)$, we would still not see significant biases from neglecting precession and higher-order multipoles in analyses. As in Sec.~\ref{sec:simulation} and Fig.~\ref{fig: impact simulated}, we see some small differences in $\beta_{q}$, with XAS preferring more symmetric binaries. Given that XP, XHM and XPHM agree well, this suggests that the inclusion of precession or higher-order multipoles is helpful for constraining the mass distribution of black holes, but not enough to provide a biased measurement. The impact of higher-order multipoles remains largely negligible, consistent with the conclusions in Sec.~\ref{sec:simulation} and~\cite{Singh:2023aqh}.

\subsection{Waveform systematics} \label{sec:model_systematics}

\begin{figure}
    \includegraphics[width=0.48\textwidth]{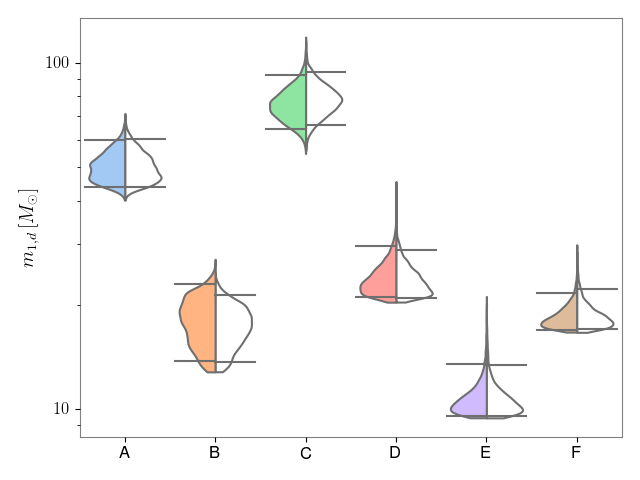}
    \includegraphics[width=0.48\textwidth]{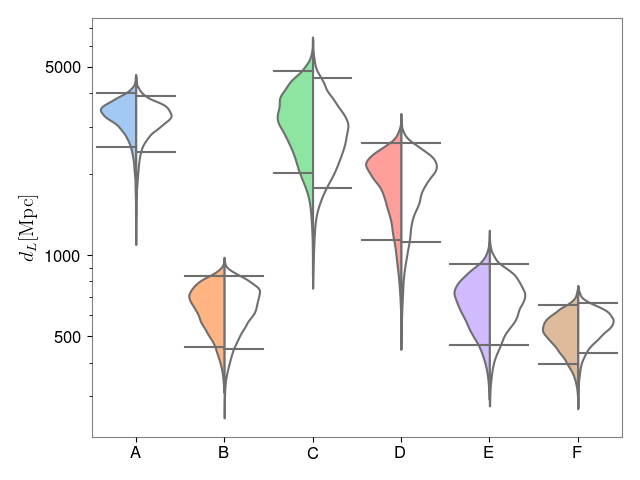}
    \caption{Violin plots showing a comparison between single-event posterior samples when analysing the same simulated signal with different models. The left panel considers the detector-frame primary mass, $m_{1, d}$, and the right panel shows the luminosity distance $d_{L}$. We randomly chose 6 signals considered in Sec.~\ref{sec:simulation}; these signals were drawn from a highly precessing population with preferentially asymmetric mass-ratios, and injected into idealised instrumental noise. The left hand side of each violin shows the result obtained with the {\texttt{SEOBNR}} waveform family ({\texttt{SEOBNRv5PHM}}) and the right hand side shows the result obtained with the {\texttt{Phenom}} waveform family ({\texttt{IMRPhenomXPHM}}, XPHM). In all cases, the simulated signals were generated with XPHM. The horizontal lines show the 90\% symmetric credible interval.}
    \label{fig:waveform_systematics}
\end{figure}

Although the primary aim of our study is to assess the impact of excluding known general relativistic phenomena (precession and higher order multipoles) on cosmological inference, and is therefore independent of the waveform family chosen, we additionally consider the impact of waveform systematics on our results; systematic errors due to mis-modelling known physical effects, which can cause discrepancies in \emph{e.g.} the inferred luminosity distance and detector-frame component masses.  This has also been explored in Refs.~\citep{Kunert:2024xqb,Dhani:2025xgt}.

We randomly chose 6 simulated signals from those considered in Sec.~\ref{sec:simulation}, due to computational limitations. These injections were drawn from a highly precessing population with preferentially asymmetric mass-ratio binaries, and injected into idealised instrumental noise. Each simulated signal was produced with XPHM, and we performed single-event Bayesian inference with the state-of-the-art model from the {\texttt{SEOBNR}} waveform family, {\texttt{SEOBNRv5PHM}}~\citep{Ramos-Buades:2023ehm}, and XPHM for comparison; similar to XPHM, {\texttt{SEOBNRv5PHM}} includes the effects of precession and higher-order multipole moments. Our injections had mass ratios varying between $0.70 \leq q \leq 0.98$, inclination angles varying between $0.39 < \theta_{\mathrm{JN}} [\mathrm{rad}] \leq 3.10$, and primary spin magnitudes varying between $0.44 \leq a_{1} \leq 0.92$. The amount of precession in the system, as described by the effective precession spin $\chi_{\mathrm{p}}$~\citep{Schmidt:2014iyl}, varied between $0.27 \leq \chi_{\mathrm{p}} \leq 0.90$, where $\chi_{\mathrm{p}} = 1$ $(0)$ indicates maximal (minimal) precession. Previous studies suggest that waveform systematics generally become increasingly significant for large $\chi_{\mathrm{p}}$, see e.g. Fig.10 in Ref.~\cite{Hamilton:2025xru} and Ref.~\cite{LIGOScientific:2025rsn}.

In Fig.~\ref{fig:waveform_systematics}, we show the marginalized single-event posterior distributions for the inferred detector-frame primary mass and luminosity distance. For all 6 cases, we see remarkable agreement between distributions, with comparable 90\% credible intervals. We see minor differences in the inferred luminosity distance for injection C, however, these differences will likely average out in population analyses; injection C was a high mass binary black hole with the lowest amount of precession out of the injections considered, $\chi_{\mathrm{p}} = 0.27$. It was also a close-to equal mass ratio binary with $q = 0.98$ inclined at $\theta_{\mathrm{JN}} = 2.84\, \mathrm{rad}$. Based on our limited results, we do not foresee waveform systematics as being a limiting factor when performing cosmological inference at current detector sensitivities. However, further injections are required to make conclusive statements, but we remind the reader that this will require significant computational resources. We leave this to future work.

\bibliographystyle{JHEP}
\bibliography{references}

\end{document}